\documentclass{aa}  
\AtBeginDocument{\let\linenumbers\nolinenumbers}

\usepackage{graphicx}
\usepackage{txfonts}
\usepackage{hyperref}

\begin{document}

\title{ALMA quiet Sun observations compared with a simulated MURaM-ChE chromosphere}

   \author{M. Loukitcheva
          \inst{1,}
          \inst{2}
         \and
          S. K. Solanki\inst{1,}\inst{3}
          \and
          D. Przybylski\inst{1}
         \and
         S.M. White
         \inst{4,}
         \inst{5}
         \and
          R. Cameron\inst{1}
          }

   \institute{Max Planck Institute for Solar System Research, Justus-von-Liebig-Weg 3, 37077 G\"ottingen, Germany\\
              \email{lukicheva@mps.mpg.de}
         \and
             Special Astrophysical Observatory, Russian Academy of Sciences (SAO RAS), Nizhniy Arkhyz, Karachay-Cherkessian Republic, 369167, Russia\\
         \and
             School of Space Research, Kyung Hee University, Yongin, Gyeonggi 17104, Republic of Korea\\
             \email{solanki@mps.mpg.de}
         \and 
             Space Vehicles Directorate, Air Force Research Laboratory, Kirtland AFB, NM, USA\\
         \and
         	Department of Physics and Astronomy, University of New Mexico, Albuquerque, NM, 87106, USA\\
                         }

   \date{Received ; accepted }

\abstract
{}
{We assess the ability of the MURaM-ChE enhanced network simulation to reproduce the chromospheric millimeter brightness observed with ALMA and investigate the properties of the simulated emission.
}
{We analyze synthetic brightness temperatures from the MURaM-ChE model in ALMA Bands~3 (3.2~mm) and 6 (1.3~mm) and compare them with ALMA observations of quiet-Sun, network, and enhanced network regions. We examine the statistical properties of the brightness distributions, their dependence on magnetic field strength, and their variation with wavelength, using results from Bifrost enhanced network snapshots as additional reference cases.}
{The MURaM-ChE model reproduces the range and variability of brightness temperatures observed in enhanced network regions, with particularly close agreement found for the 23 April 2017 data set. The reference Bifrost snapshots produce lower brightness temperatures in both bands, with values below even those measured by ALMA in the quiet Sun. The systematic offset between the models is present in both
bands and is likely associated with differences in atmospheric density
stratification and the resulting effective formation heights. Both the observations and simulations show an overall increase in brightness temperature with magnetic field strength and wavelength, although significant scatter is present.
}
{The results indicate that the MURaM-ChE model provides a physically consistent representation of chromospheric conditions in enhanced network regions and captures key properties of the observed millimeter emission. The comparison highlights the sensitivity of millimeter brightness to atmospheric stratification and magnetic structure and demonstrates the diagnostic potential of ALMA observations for constraining chromospheric models.}


   \maketitle
%

\section{Introduction}
The solar chromosphere is a crucial and still poorly understood layer in the atmosphere of the Sun that is responsible for the energy transfer between the photosphere and corona. 
Historically the solar chromosphere has been studied using strong chromospheric lines formed at wavelengths ranging from the ultraviolet to the infrared (e.g., Mg {\sc ii} h\&k, Ca {\sc ii} H\&K,  H$\alpha$, Ca {\sc ii} infrared triplet). However, these chromospheric diagnostics are not straightforward to interpret due to the fact that the lines are formed under departures from Local Thermodynamic Equilibrium (LTE) conditions, so that observed spectra cannot be directly translated into atmospheric physical properties. Moreover, the lines are non-thermally
broadened which further complicated the temperature-structure diagnostics.

An alternative diagnostic is offered by the continuum emission at millimeter (mm) wavelengths, which becomes optically thick in the chromosphere and forms under LTE conditions. In the Rayleigh–Jeans limit, mm brightness temperature and circular polarization, typically measured in radio observations,  can be directly interpreted in terms of local plasma temperature and magnetic field at the formation heights of the mm continuum. Until recently, the biggest disadvantage of mm observations was the low spatial resolution of existing mm instruments. This situation changed in 2016 with the advent of solar observations at the Atacama Large Millimeter/Submillimeter Array \citep[ALMA,][]{Wootten2009}, which is able to interferometrically map the Sun at a number of mm wavelengths with a spatial resolution of an arcsecond or less and at a cadence of 1-2 seconds \citep{Shimojo17}. ALMA contributes to a wide range of topics in solar physics, as has been demonstrated in a number of  early papers and theoretical and observational reviews \citep[see, e.g.,][]{Wedemeyer2016, Heinzel2022, Nindos2022,  Loukitcheva2022}.  

Reliable chromospheric measurements are extremely valuable for testing and validating solar atmospheric models, and the proven validity of these models will give confidence in their application to other stars as well.  Among the most widely used classical 1D semi-empirical atmospheric models are those of \citet{Vernazza1981} and \citet{Fontenla1991, Fontenla1993}. The millimeter and submillimeter signature of static and dynamic 1D models has been first studied by \citet{Loukitcheva2004}. In recent years, a new type of model has to a large extent replaced these 1D models. The new models are based on 3D radiative magnetohydrodynamic (rMHD) simulations that treat the radiative transfer of energy in non-LTE and the hydrogen atom in non-equilibrium, and cover the whole atmosphere, from the photosphere through the chromosphere to the corona. The Bifrost code \citep{Gudiksen2011, Carlsson2016} was the first to realistically implement physical processes crucial for the chromosphere in 3D. This approach produced significant advances in explaining many of the small-scale events and structures recorded by observations \citep[e.g.,][]{Carlsson2019}. The signatures produced by such models in the ALMA bands were investigated by \citet{Loukitcheva2015,
Loukitcheva2017}.

In this work we explore another representative of the 3D rMHD models, an updated version  of the MURaM
\footnote{Max-Planck-Institute for Solar System Research/University of
Chicago Radiation Magneto-hydrodynamics.}\label{muram} rMHD  code \citep{Vogler2005}, which incorporates both the coronal extension by \citet{Rempel2014, Rempel2017} and the chromospheric extension by \citet{Przybylski2022}, hereafter referred to as MURaM-ChE. The main aim of the present work is to test to what extent the MURaM-ChE simulations can reproduce the ALMA observations of the quiet Solar chromosphere in the millimeter continuum.  For this purpose we utilize a snapshot of a MURaM-ChE simulation that has a large-scale magnetic field similar to that represented by the Bifrost enhanced network (ENW) simulation from \citet{Carlsson2016}. Further details of the model are given in Sect.~\ref{model}. From the model we simulate mm continua by calculating mm brightness temperatures (Sects.~\ref{mm}, \ref{tb}), degrade them to spatial resolutions typical of solar ALMA observations, and compare them with measurements from a number of solar ALMA projects carried out in 2017-2019 (Sects.~\ref{obs}, \ref{alma},  \ref{comparison}). 

\section{Methods}\label{methods}
\subsection{3D Model of Enhanced Network}\label{model}

The MURaM-ChE code \citep{Przybylski2022} includes various extensions from the original version of \citet{Vogler2005} in order to cover the wide range of physical parameters occurring from the upper convection zone to the corona. From the coronal extension \citep{Rempel2017}, we use a slope-limited numerical diffusion scheme with high-order hyperdiffusion \citep{Rempel2014}, a hyperbolic treatment of the Spitzer heat conduction \citep{Spitzer1962}, and the "Boris"-correction \citep{Boris1970} to limit the timestep resulting from the high Alfv\'en speed in the solar corona.  In order to cover the  large range of physical conditions from the convection zone to the corona, different treatments of the transfer of energy by radiation are included, from the diffusion approximation used in sub-surface layers, to a 4-band scattering multi-group radiative transfer scheme \citep{Skartlien2000, Hayek2010} in the photosphere and above, which is combined with approximate non-LTE losses in the chromosphere based on the prescription of \citet{Carlsson2012}, while the corona is treated as an optically thin medium. The equation of state (EoS) used consists of a set of tables based on the free-EoS package \citep{Irwin2012} in the solar interior, which is smoothly merged to a non-equilibrium (NE) EoS in the photosphere and chromosphere. This NE EoS allows for a time-dependent treatment of hydrogen ionization and recombination, which is required to accurately model the solar chromosphere \citep{Sollum1999, Leenaarts2006, Leenaarts2007}. The code has been shown to reproduce with very high accuracy solar data such as the global radiative spectrum, the degree of limb darkening, and quiet Sun photospheric line profiles and bisectors \citep{Witzke2024}. 

The simulation consists of a bipolar magnetic field structure, designed to model an enhanced network region (ENW), similar to the setup used by \citet{Carlsson2016}. The simulation domain is $(24~\mathrm{Mm})^3$ with a 1024x1024x1200 grid, giving a resolution of $\approx 23.4~\mathrm{km}$ horizontally and $20~\mathrm{km}$ vertically. The vertical dimension spans from $\approx 7~\mathrm{Mm}$ below the average $\tau_{500~\mathrm{nm}}=1$ surface to $\approx 17 ~\mathrm{Mm}$ above. At the lower boundary we use the open symmetric-field condition (OSb) of \citet{Rempel2014}. The upper boundary  is open to outflows, and closed to inflows, with a potential magnetic field. The horizontal boundaries are periodic. The simulation is not an exact reproduction of the \citet{Carlsson2016} simulation: the MURaM box is deeper and additionally has approximately double the horizontal resolution. In this study, we analyze simulation snapshot 499000, which corresponds to approximately the same time after the introduction of the bipole field as snapshot 463 of the ENW Bifrost simulation presented by \citet{Carlsson2016}. The same simulation snapshot is described in detail by \citet{Ondratschek2024, Ondratschek2026}, who studied the formation of the \ion{Mg}{II} h\&k lines.
	
\begin{figure*}
   \centering
   \includegraphics[width=0.995\textwidth]{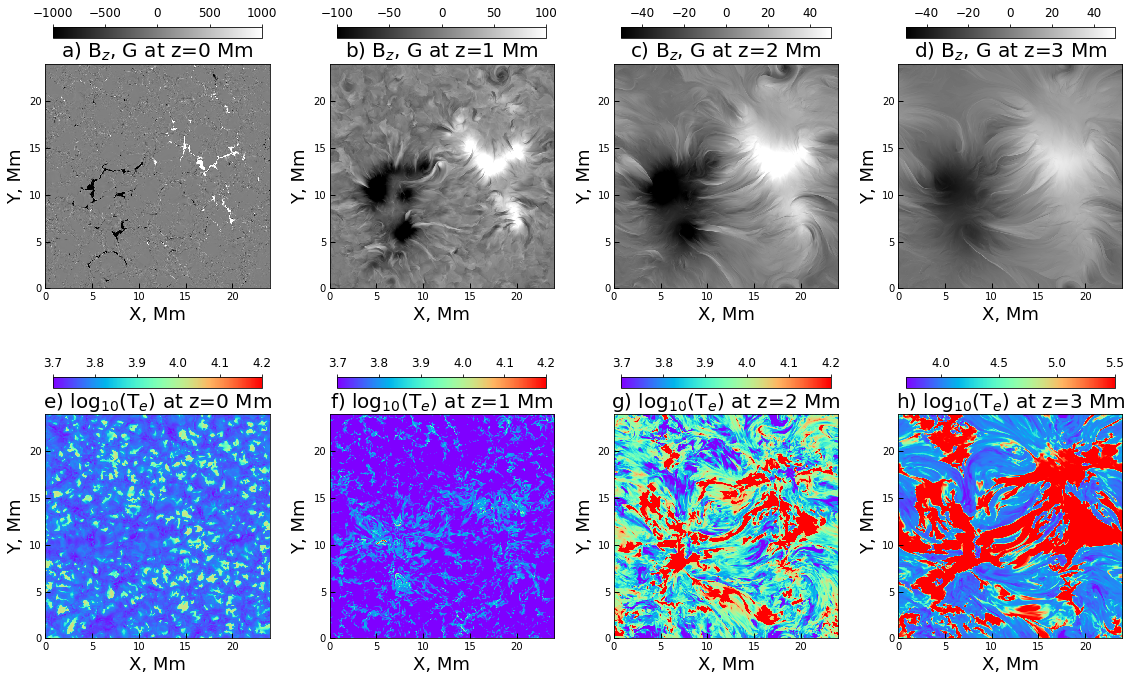}
   \caption{Vertical component of the magnetic field (panels a-d), and logarithm of electron temperature (panels e-h) at $z=0$ km, 1000 km, 2000 km and 3000 km, respectively.  The values are clipped to the limits of the corresponding color bars for visualization purposes. 
}
              \label{fig1}%
    \end{figure*}

Figure~\ref{fig1} shows the vertical component of the magnetic field  and the electron temperature at horizontal cuts through the 3D MURaM-ChE simulation at the photospheric level ($z=0$~km), and at three chromospheric heights ($z=1000, 2000, 3000$~km). These chromospheric levels roughly correspond to the formation heights of the mm continua at the frequencies of ALMA solar observations (see  Sect.~\ref{tb}). 


For the snapshot analyzed here, the mean unsigned vertical magnetic field strength is $\langle |B_z| \rangle = 64$~G at $\tau_{500}=1$, where $\tau_{500}$ denotes the continuum optical depth at 500~nm. At a $\tau_{500}=0.1$, approximately corresponding to the formation layer of the magnetic-field signal measured by the Helioseismic and Magnetic Imager \citep[HMI;][]{Scherrer2012} on board the Solar Dynamics Observatory (SDO), the horizontally averaged unsigned vertical magnetic field strength is $43.5$~G at the native model resolution and decreases to $20.7$~G after spatial degradation to the $\sim$1\arcsec\ resolution typical of SDO/HMI \citep{Ondratschek2024}.


\subsection{Brightness at millimeter wavelengths}\label{mm}

The formation of the millimeter/submillimeter continum in the quiet solar atmosphere is well understood and has been thoroughly discussed in numerous studies, e.g. \citet{Zheleznyakov1968, Dulk1985, Loukitcheva2004, Wedemeyer2016}. In this work we follow the approach of \citet{Loukitcheva2015}, who analyzed the millimeter continua from the Bifrost ENW simulation.  From the 3D MURaM-ChE ENW snapshot we synthesize the mm/submm brightness at wavelengths from 0.4 mm to 8.6 mm, which correspond to the full wavelength coverage of the ALMA interferometer,  including the frequency bands that ALMA commonly offers for solar observations, Bands 6 and 3, as well as the highest and lowest frequency bands to be incorporated in the future. 
The emergent millimeter continuum intensity is calculated by formally integrating the radiative transfer equation along each vertical line of sight through the simulation snapshot. The opacity includes thermal free-free contributions from electron--ion and electron--neutral hydrogen interactions \citep[see][for details]{Loukitcheva2004, Loukitcheva2015}.

\subsection{ALMA observations}\label{obs}

Since the start of regular solar observations with ALMA in 2016, a limited number of data sets has been accumulated in Bands 6 and 3. Historically, Band 3, ranging from 84 to 116 GHz (wavelengths of 2.6-3.6~mm), has been the primary frequency range for millimeter studies of the solar chromosphere, even prior to ALMA, because it combines relatively moderate atmospheric requirements with a formation height in the middle chromosphere that allows comparatively straightforward physical interpretation. As a result, there are more ALMA observations in Band 3 than in Band 6, which covers frequencies from 211 to 275 GHz (wavelengths of 1.1-1.4~mm). Band 6 is more demanding in terms of weather conditions due to the atmospheric response at higher observing frequencies and typically involves more complex formation lower in the solar chromosphere (e.g. wide and double-peaked forms of intensity contribution functions, \citet{Loukitcheva2015}. 

Table~\ref{table_observations} provides a summary of ALMA data sets used in this study for comparison with the ENW models. For each data set we list the observation date, ALMA project code, observing band, mean brightness temperature with its standard deviation, the average unsigned photospheric magnetic field strength derived from co-spatial SDO/HMI magnetograms, the type of the region observed, and the data origin (literature reference or this work). The type of the observed region was determined based on the value and morphology of the photospheric magnetic field, including field extrapolations to chromospheric heights \citep[where available,][]{Jafarzadeh2021}. We distinguish the following target types for the data sets considered in Table~\ref{table_observations}: qs - quiet Sun; enw - enhanced network; nw - network; pl - plage; and pore). Additionally, for data sets with pronounced horizontal fields within FOVs we add ''+hor'' to the target type. The temperature statistics listed in Table~\ref{table_observations} were derived over the full available fields of view. For the data sets examined in detail in Sect.~\ref{alma}, Tables~\ref{tb_stats_band3} and ~\ref{tb_stats_band6} report values for the selected subregions and therefore may differ from those in Table~\ref{table_observations}.

We note that the data sets used in this work and listed in Table~\ref{table_observations} do not represent the complete set of ALMA solar observations obtained to date. Rather, they focus on early observations for which reliable brightness measurements have been published \citep{Jafarzadeh2021, Nindos2018}, as well as data products that were made publicly available in a form suitable for analysis  \citep{Kobelski2022} or provided by the authors \citep{Molnar2019, Molnar2021}, and also observations from projects in which the authors participated directly \citep[e.g.][] {Loukitcheva2019}.

Data sets labeled J2021 in Table~\ref{table_observations} correspond to the observations presented by \citet{Jafarzadeh2021}, while data sets 2-6, labeled N2018, refer to quiet-Sun observations reported by \citet{Nindos2018}. For these data sets, the brightness temperatures are taken directly from the respective publications. The corresponding photospheric magnetic field values are derived from HMI magnetograms using the reported positions and spatial extents of the ALMA fields of view (FOVs). For the J2021 observations, the ALMA FOVs span approximately $46$–$52\arcsec$ in Band~3 and $22$–$25\arcsec$ in Band~6, while for the N2018 observations the Band~3 FOV extends up to about $60\arcsec$.

Eight data sets, numbered 7, 8, 10, 12, 14, 15, 18, and 21, including the ones labeled K2021, M2019, M2021, and ``this work'', are used for a detailed comparison between model and observed brightness distributions and are analyzed in Sect.~\ref{alma}. 

It is important to note that the ALMA interferometer observations presented here measure spatial brightness-temperature variations but do not recover the absolute background level because they lack zero-spacing information. To establish an absolute brightness-temperature scale, a background offset is determined from simultaneous single-dish measurements, which provide absolute brightness temperatures at much poorer spatial resolution. To establish a repeatable scale at about the 100~K uncertainty level, \citet{White2017} recommended that disk-center quiet-Sun temperatures of 7300~K at Band 3 and 5900~K at Band 6 be enforced in the single-dish maps, based on a large set of commissioning data calibrated from first principles and relying on (possibly imperfect) knowledge of telescope aperture efficiencies. \citet{Alissandrakis2017, Alissandrakis2020} took a different approach based on the scaling of opacity with frequency, finding that 7300~K is appropriate at Band 3 but arguing that Band 6 should be about 6300~K in the quiet Sun at disk center. Thus there may be uncertainty in the base level of Band 6 temperatures used here of up to a few hundred K.

\begin{table*}[h!]
\centering
\begin{tabular}{llllllll}
\hline
\textbf{Data set} & \textbf{Date} & \textbf{Project} & \textbf{Band} & \textbf{$T_b$ (K)} & \textbf{|$B_z$| (G)} & \textbf{Target}  & \textbf{Origin} \\
\hline
1 & 2016-12-22 & 2016.1.00423.S & 3 & 7470 $\pm$ 480 & 11.7 & qs & J2021 \\
2 & 2017-03-16 & 2016.1.00572.S & 3 & 7460 $\pm$ 440 & 9.7 & qs & N2018 \\
3 & 2017-03-16 & 2016.1.00572.S & 3 & 7270 $\pm$ 400 & 10.0 & qs & N2018 \\
4 & 2017-03-16 & 2016.1.00572.S & 3 & 7330 $\pm$ 430 & 10.2 & qs & N2018\\
5 & 2017-03-16 & 2016.1.00572.S & 3 & 7270 $\pm$ 460 & 9.7 & qs & N2018\\
6 & 2017-03-16 & 2016.1.00572.S & 3 & 7220 $\pm$ 390 & 9.9 & qs & N2018\\
7 & 2017-03-19 & 2016.1.00030.S & 3 & 7465 $\pm$ 570 & 33.7 & enw & this work \\
8 & 2017-03-21 & 2016.1.00788.S & 3 & 7445 $\pm$ 425 & 31.0 & enw & K2021, this work \\
9 & 2017-04-22 & 2016.1.00050.S & 3 & 9465 $\pm$ 1129 & 97.8 & pl & J2021 \\
10 & 2017-04-23 & 2016.1.01129.S & 3 & 9065 $\pm$ 1225 & 20.0 & enw & M2019, this work\\
11 & 2017-04-27 & 2016.1.01532.S & 3 & 7728 $\pm$1150 & 54.1 & nw+hor & J2021\\
12 & 2017-04-27 & 2016.1.00202.S & 3 & 7575 $\pm$ 855 & 16.0 & nw & L2019, this work \\
13 & 2018-04-12 & 2017.1.00653.S & 3 & 7663 $\pm$ 625 & 9.5 & qs & J2021 \\
14 & 2019-11-09 & 2019.1.00516.S & 3 & 7315 $\pm$ 490 & 10.0 & qs & this work \\
\hline
15 & 2017-04-16 & 2016.1.00202.S & 6 &5925 $\pm$ 780  & 8.5 & qs & this work\\
16 & 2017-04-18 & 2016.1.01129.S & 6 & 7324 $\pm$ 954  & 35.8 & nw+hor  & J2021\\
17 & 2017-04-22 & 2016.1.00050.S & 6 &7746 $\pm$ 859  & 149.3 & pl & J2021 \\
18 & 2017-04-23 & 2016.1.01129.S & 6 & 6940 $\pm$ 925 & 24.0 & enw & M2021,  this work\\
19 & 2018-04-12 & 2017.1.00653.S & 6 &5957 $\pm$ 352  & 8.4 & qs & J2021\\
20 & 2018-08-23 & 2017.1.01672.S & 6 &6104 $\pm$ 497  & 30.5 & pore+hor& J2021 \\
21 & 2019-11-09 & 2019.1.00516.S & 6 &5905 $\pm$ 415  & 10 & qs  & this work\\
\hline
\end{tabular}
\caption{Summary of ALMA data sets analyzed in this study. The brightness-temperature statistics refer to the full available fields of view. See the text for definitions of the target classifications. The label J2021 refers to \citet{Jafarzadeh2021}, N2018  - \citet{Nindos2018}, K2021 - \citet{Kobelski2022}, L2019 - \citet{Loukitcheva2019}, M2019 and M2021- \citet{Molnar2019, Molnar2021}, respectively. The horizontal line separates data sets recorded in Bands 3 and 6.}
\label{table_observations}
\end{table*}

\section{Results and Discussion}\label{res}

\subsection{Simulated millimeter brightness}\label{tb} 

\begin{figure*}[!h]
   \centering
   \includegraphics[width=0.995\textwidth]{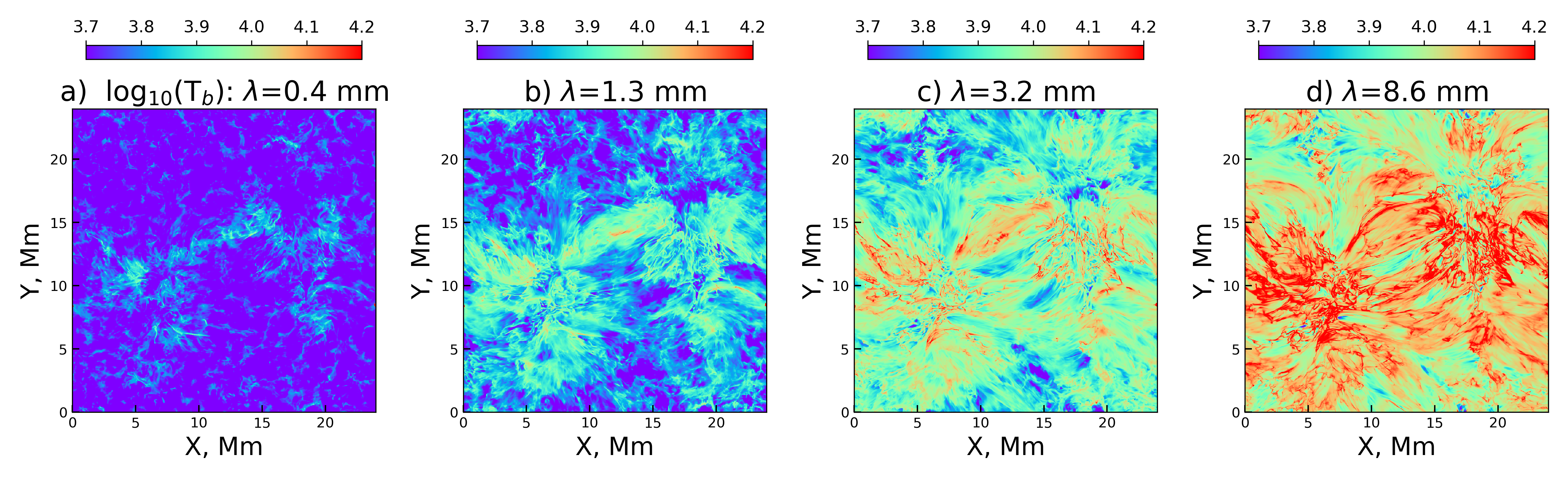}
   \caption{Simulated brightness temperatures at the resolution of the model for the analyzed wavelengths.  The values
 are clipped to the limits of the corresponding color bars for visualization purposes. 
}
              \label{fig2}
    \end{figure*}

Figure~\ref{fig2} shows the results of the mm brightness simulations at four wavelengths corresponding to ALMA Bands 6 and 3 (1.3 mm and 3.2 mm, respectively), as well as to the shortest and longest wavelengths that may potentially be used in future solar observations (0.4 mm and 8.6 mm). The corresponding minimum and maximum brightness temperatures, together with the mean values and effective formation heights averaged over the computational domain, are listed in Table~\ref{table1}. For each wavelength, the effective formation height, $h_{eff}$, was calculated as the average of the centroids of the corresponding intensity contribution functions \citep[see][for details]{ Loukitcheva2004, Loukitcheva2015}. We will further refer to these heights as the approximate heights of formation of mm emission in the analyzed model snapshot. A detailed discussion of formation heights in a dynamic atmosphere is beyond the scope of the present study.

The brightness distribution at ALMA wavelengths exhibits structures that are visually well correlated with the magnetic field at chromospheric heights. Thus, already at $z=1000$~km small magnetic loops are visible in the vicinity of the two main magnetic concentrations (Fig.~\ref{fig1}b). This behavior is reproduced in the brightness distribution at 0.4~mm (Fig.~\ref{fig2}a), where elongated structures of enhanced brightness, resembling small magnetic arches between opposite-polarity concentrations, are seen in the same regions. Interestingly, the fibrilar pattern is less pronounced in the electron temperature distribution at the height of $1000$~km (Fig.~\ref{fig1}f), suggesting that already at short mm wavelengths the mm brightness reflects the magnetic topology at the corresponding effective formation heights. For easier visual comparison between mm brightness and electron temperature at the respective heights of formation, we used the same display range of $[3.7,4.2]$ for the logarithm of temperature in Figs.~\ref{fig1} and \ref{fig2}, except for the electron temperature at $z=3000$~km, which reaches significantly higher values.

   \begin{figure*}[!h]
   \centering
   \includegraphics[width=0.96\textwidth]{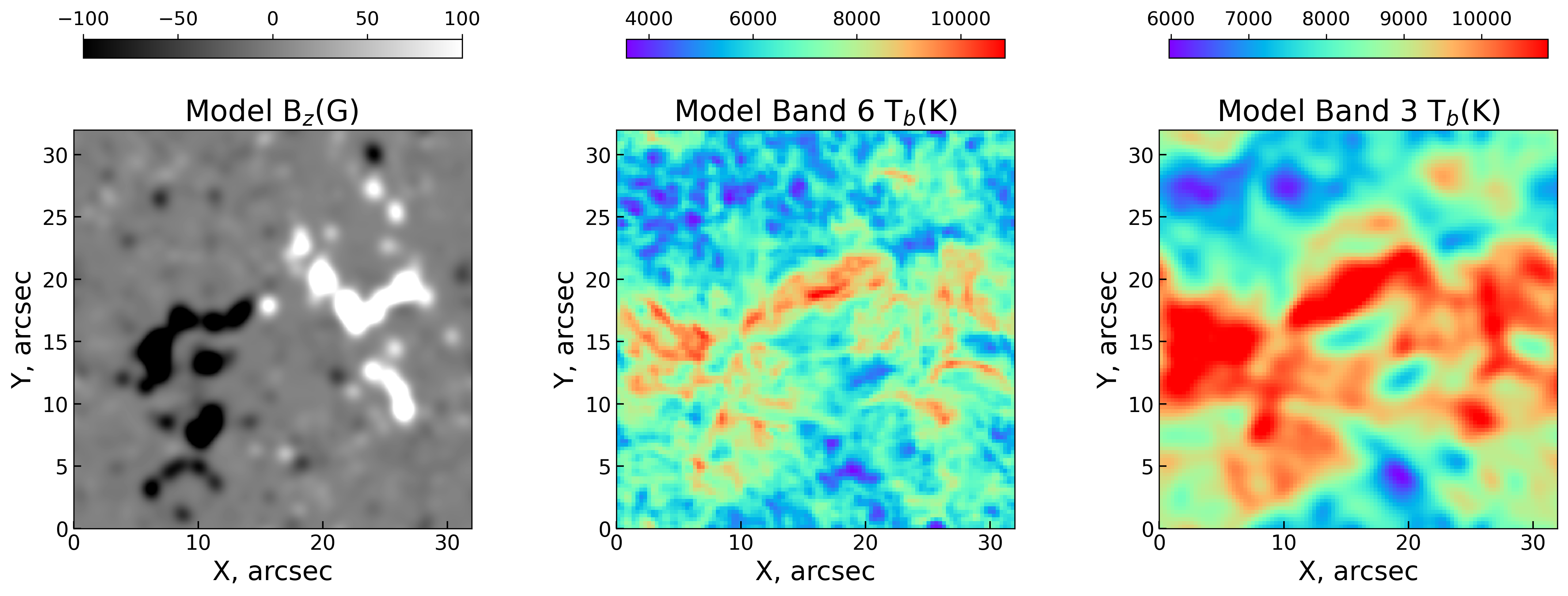}
   \caption{Vertical component of the model magnetic field evaluated at the HMI formation height, together with simulated brightness temperature maps at ALMA Band 6 (1.3 mm) and Band 3 (3.2 mm), degraded to the ALMA spatial resolution of 1\arcsec.  The values
 are clipped to the limits of the corresponding color bars for visualization purposes.  
}
              \label{fig_bz_tb}
    \end{figure*}

Higher in the chromosphere, at $z=2000$~km and $z=3000$~km, the magnetic field becomes increasingly inclined and magnetic loops grow more prominent within the field of view, connecting also the main magnetic concentrations (Fig.~\ref{fig1}c–d). A similar morphological evolution is seen in the mm brightness at longer wavelengths (Figs.~\ref{fig2}b–d), consistent with the increase in effective formation height. With increasing wavelength, the fibrilar structures become brighter, more extended, and more horizontally organized. At 8.6~mm (Fig.~\ref{fig2}d), horizontal fibrils fill a substantial fraction of the FOV.

The upper left quadrant of the simulation contains no strong magnetic concentrations and resembles a quiet-Sun internetwork region (Fig.~\ref{fig1}). In this region, shock-like patterns are expected to form as a result of the interaction of wave fronts propagating from different directions. This effect is most clearly seen in the brightness at 1.3~mm (Fig.~\ref{fig2}b), which forms at an average height of approximately $2000$~km (see Table~\ref{table1}). At the shorter wavelength of 0.4~mm (Fig.~\ref{fig2}a; formation height $h_{eff} \sim1000$~km), the shocks are not yet sufficiently developed to produce a pronounced pattern, while at wavelengths longer than 1.3~mm (with formation heights approaching $\sim3000$~km), extended horizontal structures dominate and partly obscure the shock-like signatures (Figs.~\ref{fig2}c–d).

\begin{table*}
\caption{Minimum, maximum and mean brightness temperature, together with mean effective formation height, listed at the spatial resolution of the model, and mean brightness temperature at a resolution of 1\arcsec,  for four analyzed wavelengths}             
\label{table1}      
\centering                          
\begin{tabular}{l c c c c c}        
\hline\hline                 
\noalign{\smallskip}
$\lambda$, mm & $T_{b}^{min}$, K & $T_{b}^{max}$, K & $T_b$, K & $h_{eff}, km$ & $T_b(@1'')$, K\\    
\noalign{\smallskip}
\hline                        
\noalign{\smallskip}
0.4 & 3000      &   12000 & 5000$\pm$830  &1000$\pm$270 &5000$\pm550$ \\
1.3 & 2800 & 15500 & 6800$\pm$1400  &1900$\pm$550 &6800$\pm1200$\\
3.2 & 2500   &     24550 & 8950 $ \pm$1650     & 2700$\pm$850 &8940$\pm1150$\\
8.6 & 3350  &    55650  & 11400$\pm$2400 & 3300$\pm$1100 &11400$\pm1500$ \\
\noalign{\smallskip}
\hline                                   
\end{tabular}
\end{table*}

 \begin{figure}
   \centering
    \includegraphics[width=\columnwidth]{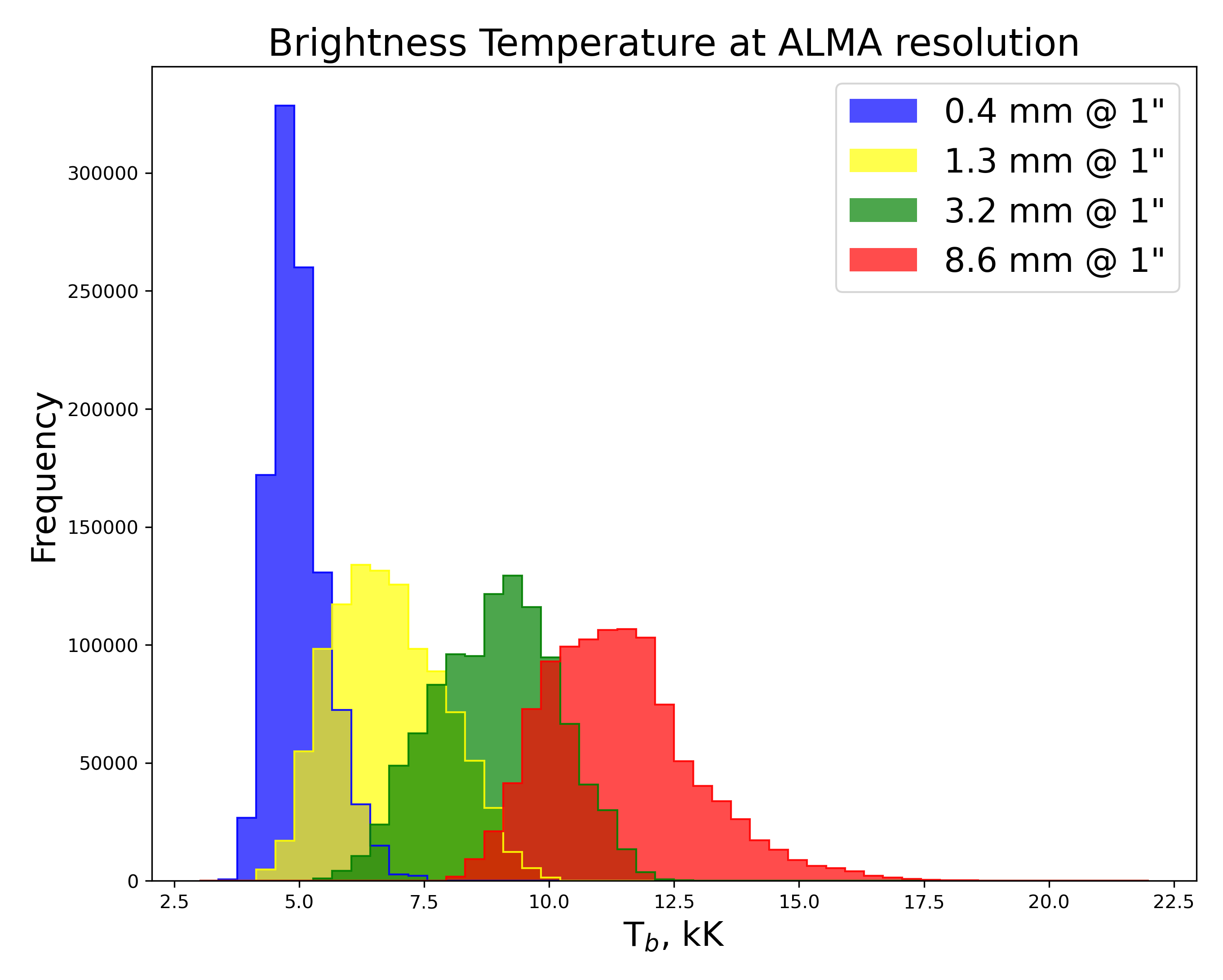}
   \caption{Distribution of brightness temperature at a spatial resolution of 1\arcsec, approximately corresponding to the resolution of the available ALMA observations, at  the  wavelengths of $0.4$~mm, $1.3$~mm, $3.2$~mm, and $8.6$~mm. }
              \label{fig3}%
    \end{figure}
To facilitate comparison with the ALMA observations, we degraded the simulated brightness from the MURaM-ChE model to a spatial resolution of 1\arcsec, comparable to the typical resolution achieved in early ALMA solar observations in Bands 6 and 3. To this end, the simulated 2D brightness images were convolved with a Gaussian kernel of 1\arcsec\ full width at half maximum (FWHM). The resulting degraded brightness temperatures are shown in Fig.~\ref{fig_bz_tb} for Bands 6 and 3 in the form of 2D maps together with the corresponding vertical magnetic field maps, mimicking HMI magnetograms, and in Fig.~\ref{fig3} as 1D brightness distributions at four wavelengths.

The spatial degradation visibly suppresses fine scale structure and reduces the contrast of the brightness maps. In particular, the small scale fibrilar and shock like features seen at the original model resolution become smoother and more spatially coherent, while the large scale brightness enhancements associated with magnetic concentrations remain clearly visible (Figs.~\ref{fig2}b-c and ~\ref{fig_bz_tb} for 1.3~mm (Band 6) and 3.2~mm (Band 3), respectively) . The corresponding mean brightness temperatures $T_b(@1'')$ are listed in the last column of Table~\ref{table1}. As expected, spatial smearing decreases the scatter and narrows the range of simulated values relative to the original values, while leaving the mean brightness levels essentially unchanged.

 The brightness distributions at different wavelengths overlap substantially (Fig.~\ref{fig3}), although systematic differences are evident. The brightness at 0.4~mm shows the smallest scatter around the mean (blue histogram), consistent with its formation in lower chromospheric layers where temperature fluctuations are comparatively modest after spatial averaging. The distribution at 8.6~mm is the broadest among the four wavelengths and exhibits a pronounced high-temperature tail (red histogram), reflecting the contribution of hotter and more spatially extended structures in the upper chromosphere. The intermediate wavelengths (1.3 and 3.2~mm) show increasing width and asymmetry.  The mean brightness temperature $T_b$ in the mm range increases with wavelength as the effective formation heights shift to higher, on average hotter, layers of the chromosphere (see Figs.~\ref{fig2}, \ref{fig3}, and Table~\ref{table1}). This monotonic trend, together with the wavelength-dependent change in the shape of the brightness distributions, illustrates the tomographic capability of multiwavelength mm observations to probe the vertical thermal structure of the chromosphere at ALMA resolution.
 
For comparison with the ALMA observations, presented in Sect.~\ref{alma}, we further focus on the simulated brightness in ALMA Bands~6 and 3, corresponding to wavelengths of 1.3~mm and 3.2~mm, respectively. The statistical properties of the simulated brightness are summarized in Tables~\ref{tb_stats_band3} and \ref{tb_stats_band6}, together with the observational values.

For reference, these tables also include brightness distributions derived from the Bifrost snapshots 385 and 463. The Bifrost snapshot 385 is used as a benchmark case \citep{Carlsson2016}, whereas snapshot 463 provides the most directly comparable Bifrost case to the MURaM-ChE snapshot analyzed here.

\subsection{Observed ALMA brightness}\label{alma}

\subsubsection{Band 3}\label{band3}

For a detailed comparison between modeled and observed brightness distributions in ALMA Band 3, we analyze five observational data sets summarized in Table~\ref{tb_stats_band3} together with the models. With one exception, all selected regions correspond to network (NW) or enhanced-network (ENW) targets and therefore represent appropriate observational counterparts to the ENW MURaM-ChE simulation snapshot. Prior to analysis, all ALMA data were spatially co-aligned with SDO/HMI magnetograms and SDO/AIA images to correct for coordinate uncertainties of up to a few arcseconds, which are common in solar interferometric observations. 

For each data set, which typically covers approximately
$60\arcsec \times 60\arcsec$, we extracted a
$30\arcsec \times 30\arcsec$ (or $35\arcsec \times 35\arcsec$)
subregion containing representative magnetic structures. This spatial extent closely matches the horizontal size of the model computational domain, ensuring meaningful comparison between simulations and observations while providing statistically robust brightness distributions.

\begin{figure}[!h]
   \centering
    \includegraphics[width=\columnwidth]{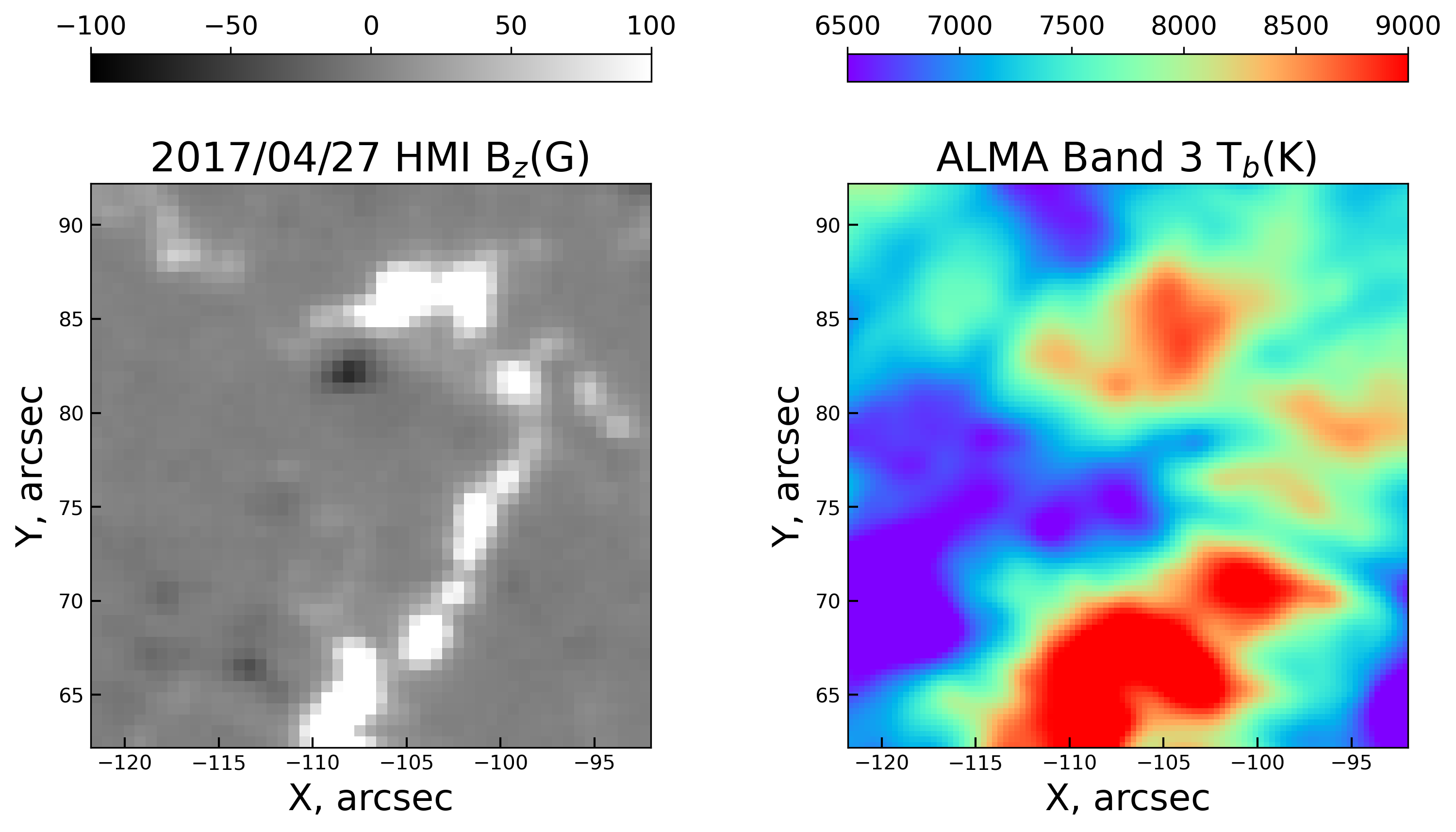}
    \caption{Time-averaged SDO/HMI vertical magnetic field (left) and ALMA Band 3 brightness temperature (right) for the 2017 April 27 observations. The averaging is performed over the ALMA observing interval. The values are clipped to the limits of the corresponding color bars for visualization purposes. }
              \label{fig_04272017}%
    \end{figure}

\textbf{April 27, 2017}:

The interferometric single-pointing observations of a quiet-Sun network region were acquired on 27 April 2017 during 16:00-16:44 UT, with a spatial resolution of approximately $1.6\arcsec$. The data cube consists of 1090 snapshots obtained with a cadence of 2 s and mapped onto a grid with a pixel size of $0.3\arcsec$. A detailed description of the observations and data analysis is provided by \citet{Loukitcheva2019}.

The selected region is centered on a network structure dominated by positive polarity, forming a characteristic question-mark-shaped configuration surrounded by a few smaller opposite-polarity patches (Fig. ~\ref{fig_04272017}). The time-averaged magnetogram shows a magnetic field range of -65 G to 350 G (clipped to the range (-100,100)~G for visualization purposes in Fig. ~\ref{fig_04272017}) and a mean unsigned field strength of approximately 18 G, comparable to the MURaM-ChE snapshot, although somewhat weaker.

The corresponding ALMA Band 3 brightness map exhibits spatial structuring closely aligned with the magnetic topology, with enhanced brightness co-spatial with stronger magnetic concentrations and lower brightness in weak-field regions. The brightness distribution has a mean temperature of 7565 K, a standard deviation of 825 K, and spans from about 5435 K to 10 385 K (Table ~\ref{tb_stats_band3}). The range of the observed brightness values is comparable with those expected from the MURaM-ChE snapshot but the mean observed temperature is substantially lower.

\begin{table*}
\centering
\caption{Statistical properties of the brightness temperature distributions in ALMA Band~3 for the selected observational subregions
and simulations analyzed in this work. The mean, median, standard deviation, minimum, and maximum brightness temperatures are given in K and rounded to the nearest 5~K. }
\label{tb_stats_band3}
\begin{tabular}{lrrrrrrr}
\hline
Data set & Mean & Median & Std & Min & Max & Skewness & Kurtosis \\
\hline
2017-03-19   & 7330 & 7370 & 505  & 5990 & 8760  & -0.16 & -0.21 \\
2017-03-21   & 7460 & 7480 & 450  & 6385 & 8810  & -0.11 & -0.40 \\
2017-04-23   & 9070 & 9130 & 1235 & 6410 & 11970 &  0.04 & -0.77 \\
2017-04-27   & 7565 & 7475 & 825  & 5435 & 10385 &  0.39 & -0.04 \\
2019-11-09   & 7295 & 7280 & 455  & 5810 & 9680  &  0.31 &  1.11 \\
\hline
MURAM        & 8940 & 9020 & 1150 & 5880 & 11720 & -0.13 & -0.61 \\
Bifrost 385  & 6205 & 6015 & 980  & 3920 & 10415 &  1.08 &  1.33 \\
Bifrost 463  & 5650 & 5480 & 1045 & 3455 & 9630  &  1.13 &  1.46 \\
\hline

\end{tabular}
\end{table*}

\begin{figure}[!h]
   \centering
    \includegraphics[width=\columnwidth]{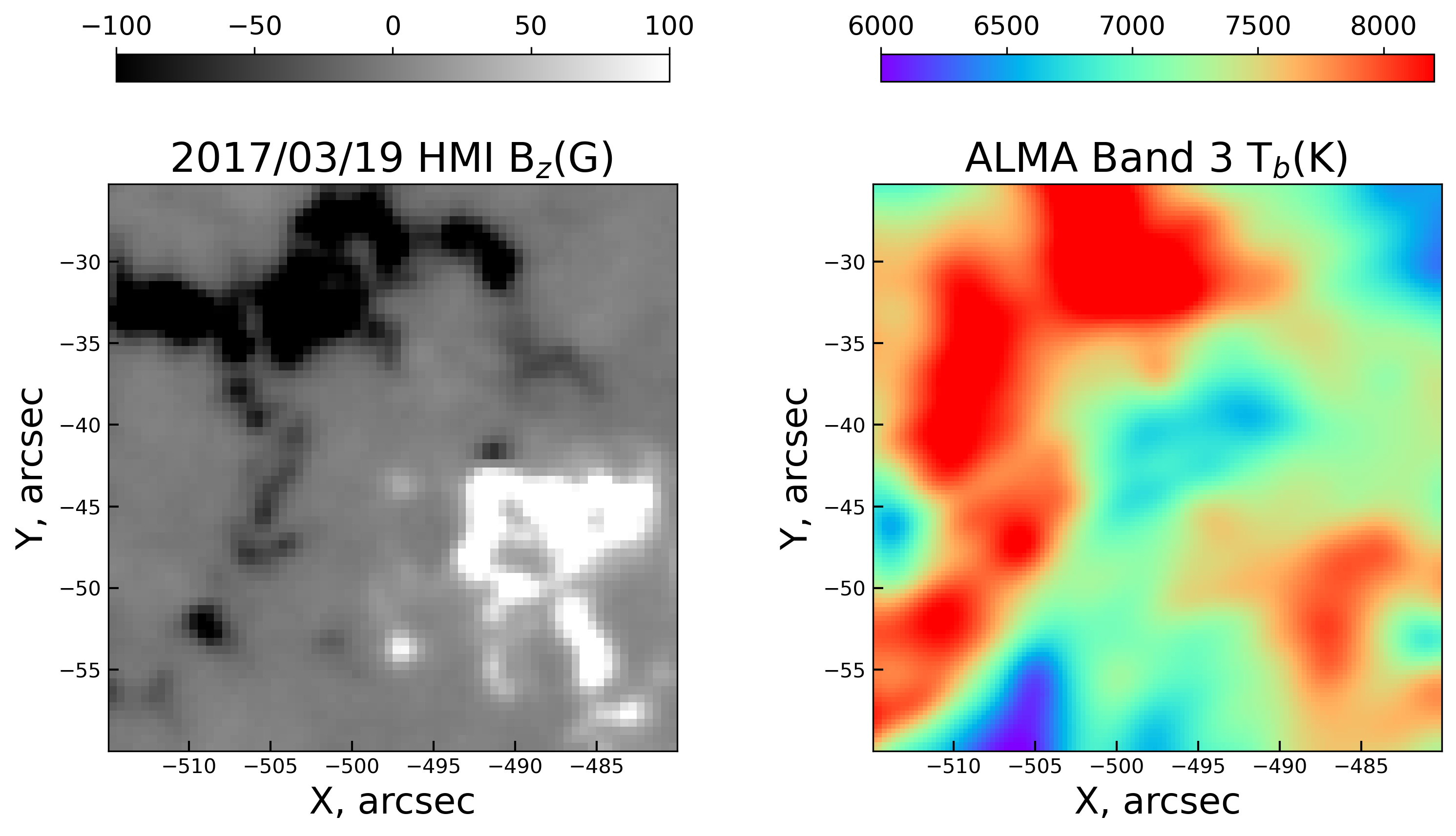}
    \caption{Same as in Fig.~\ref{fig_04272017} for the 2017 March 19 observations. 
    }
              \label{fig_03192017}%
    \end{figure}

\textbf{March 19, 2017}:

The observations obtained on 19 March 2017 during 15:32-17:40 UT sample a magnetically active enhanced-network region (Fig.~\ref{fig_03192017}). The data cube contains 2550 snapshots with a cadence of 2~s and a pixel size of $0.4\arcsec$. The observations were performed in the C-1 array configuration, with the images restored using a $2\arcsec \times 2\arcsec$ beam. Since no total-power data were available, an offset of 7300~K was applied to set the absolute brightness-temperature scale.  

The selected region exhibits strong magnetic structuring, with vertical field strengths ranging from $-557$ G to $+404$ G and a mean unsigned field of approximately 34 G, exceeding that of the ENW model snapshot. The brightness temperature varies between approximately 5990 K and 8760 K, with a mean of 7330 K and a standard deviation of 505 K (Table~\ref{tb_stats_band3}). 

Compared to the 27 April 2017 data set described above, the 19 March 2017 region exhibits stronger peak magnetic field values but a somewhat narrower brightness distribution, as reflected by the smaller standard deviation listed in Table~\ref{tb_stats_band3}.

\textbf{March 21, 2017}

The quiet-Sun region observed on March 21, 2017 was previously analyzed by \citet{Kobelski2022}, and the observational data sets were made publicly available. The ALMA observations were acquired during the time interval 15:42-16:45~UT. The data were mapped with a pixel size of $0.75\arcsec$. 

\begin{figure}[!h]
   \centering
    \includegraphics[width=\columnwidth]{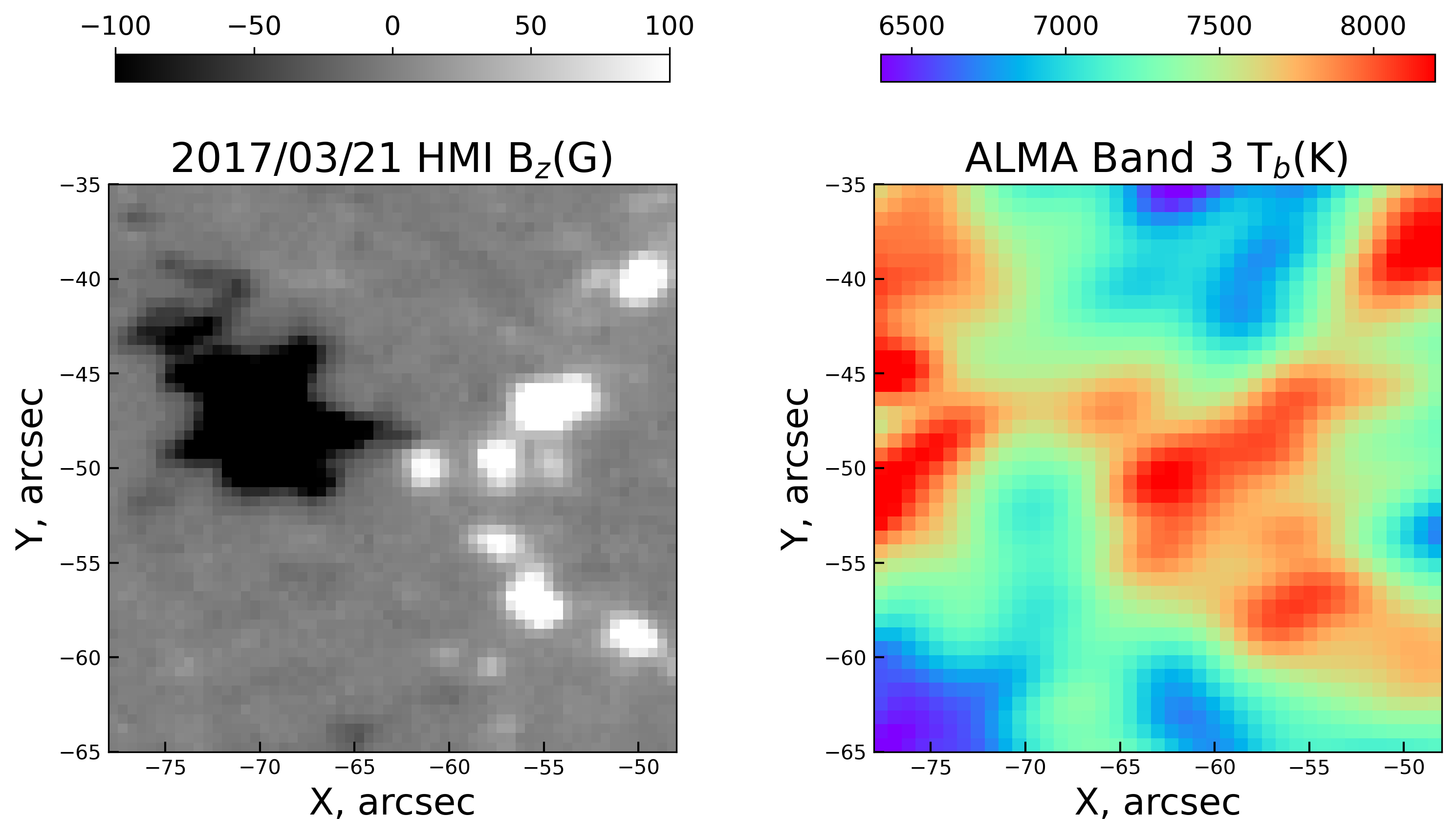}
    \caption{Same as in Fig.~\ref{fig_04272017} for the 2017 March 21 observations. }
              \label{fig_03212017}%
    \end{figure}

The 30\arcsec x 30\arcsec\  region selected for analysis (Fig.~\ref{fig_03212017}) exhibits a well developed network configuration characterized by an extended negative polarity concentration occupying much of the left portion of the FOV, while positive polarity elements are distributed more sparsely across the right side of the region. The mean unsigned magnetic field strength is approximately 31~G, which is about $30\%$ higher than in the MURaM-ChE snapshot.

The brightness distribution is characterized by a mean temperature of 7460 K and a standard deviation of 450 K (Table~\ref{tb_stats_band3}). The relatively small scatter, despite the presence of substantial magnetic structuring, is likely influenced by the coarse pixel size and relatively low spatial resolution, which reduces the apparent amplitude of small scale brightness enhancements. The observed brightness levels are comparable to the March 19 2017 data set.

\textbf{April 23, 2017}

\begin{figure}[!h]
   \centering
    \includegraphics[width=\columnwidth]{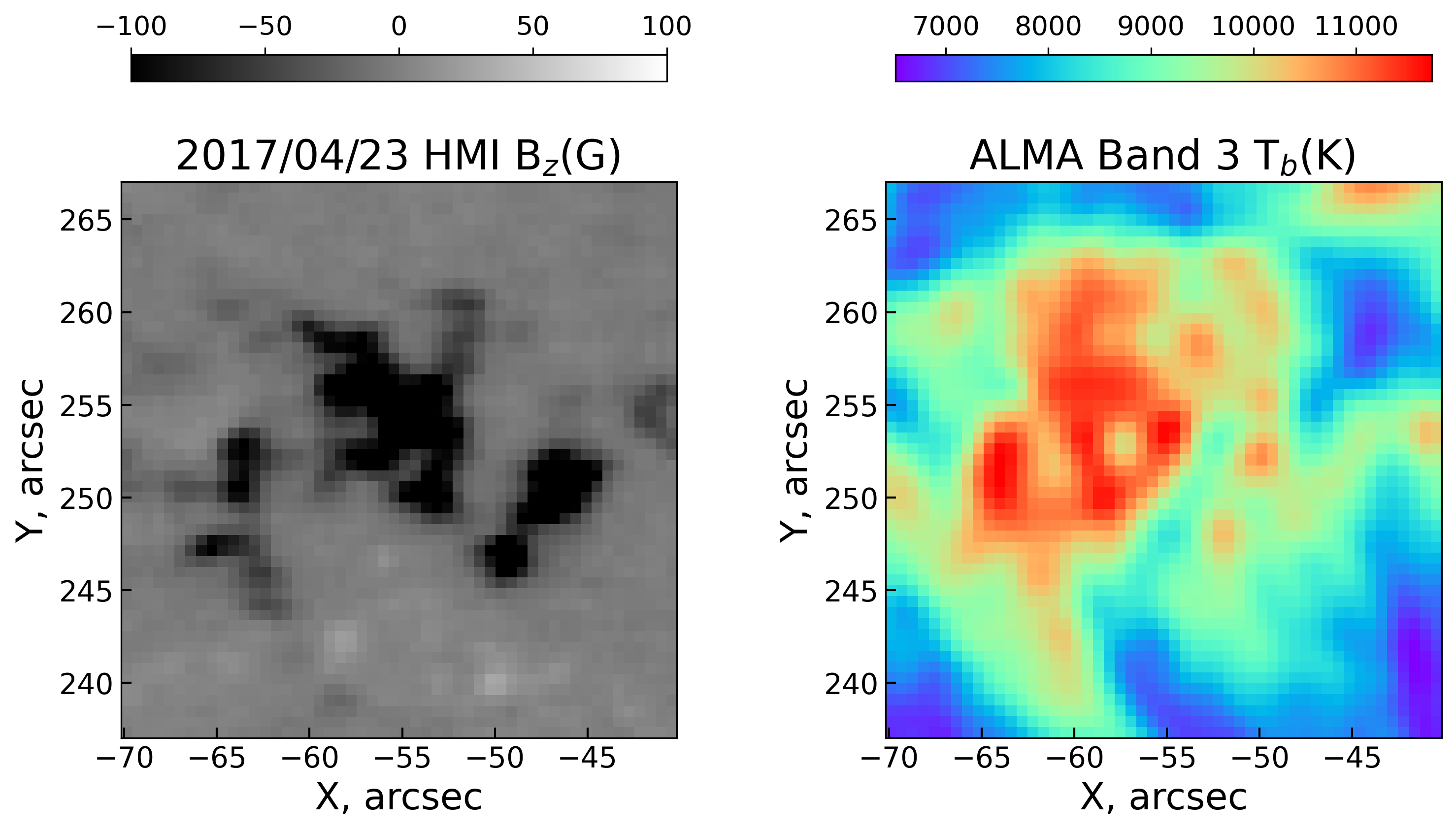}
    \caption{Same as in Fig.~\ref{fig_04272017} for the 2017 April 23 observations. }
              \label{fig_04232017_b3}%
    \end{figure}

The observations, originally presented and analyzed by 
\citet{Molnar2019,Molnar2021}, were obtained in the vicinity of a 
sunspot and plage and include compact magnetic concentrations within 
the ALMA field of view (Fig.~\ref{fig_04232017_b3}). The brightness-temperature maps were synthesized with a beam of approximately 
$1.75\arcsec \times 1.91\arcsec$ and sampled with a pixel size of 
$0.6\arcsec$. The processed maps, together with the Band~6 data 
discussed below, were provided by the authors for use in this study.

The selected subregion, located away from the plage, and observed during the interval 17:19-17:29~UT, contains clustered enhanced-network concentrations dominated by negative polarity, with peak field strengths reaching $-322$~G and a mean unsigned field strength of about 20~G (Fig.~\ref{fig_04232017_b3}). The magnetic field is organized in compact and spatially clustered patches, producing a strongly asymmetric configuration within the FOV.

The corresponding Band~3 brightness map exhibits pronounced spatial structuring closely aligned with the magnetic topology. The highest brightness temperatures are nearly co-spatial with the strongest negative-polarity concentrations. The extended central magnetic patch is associated with a broad region of enhanced mm emission. 
The brightness temperature distribution exhibits the highest mean value and largest variability among the analyzed NW and ENW data sets, ranging from approximately 6410~K to 11970~K, with a mean of 9070~K and a standard deviation of 1235~K (Table~\ref{tb_stats_band3}).

\textbf{November 09, 2019}
 
 A single-pointing observation nominally targeting a quiet-Sun region was obtained on 9 November 2019 during 15:44-17:10 UT. The images were primary-beam corrected and restored with a synthesized beam of approximately $1.2\arcsec$ and a pixel size of $0.2\arcsec$. During this observation the telescope control software applied an inappropriate gravitational deflection correction, which resulted in uncertainty of  the exact pointing location at the arcminute level, so precise coalignment with other instruments is limited.

This data set represents the magnetically quietest region among those analyzed, with unsigned magnetic field strengths not exceeding approximately 10 G. Consequently, it provides a useful reference for low activity chromospheric conditions, although it is less suitable for direct comparison with the ENW snapshot. The brightness temperature distribution ranges from about 5810 K to 9680 K, with a mean of 7295 K and a standard deviation of 455 K (Table \ref{tb_stats_band3}). The relatively narrow distribution and reduced variability reflect the absence of strong magnetic concentrations capable of producing significant chromospheric heating.

\subsubsection{Band 6}\label{band6}

As noted earlier, Band 6 solar observations are less numerous than those in Band 3, primarily due to more stringent atmospheric requirements, and the smaller effective FOV, which is typically limited to about 25\arcsec. The smaller spatial coverage also makes coalignment with complementary instruments more challenging, since fewer large-scale magnetic structures are available as reference features. For this study, we analyzed three Band 6 data sets, including one representing ENW conditions and two corresponding to very quiet-Sun regions. The statistical properties of the brightness distributions are summarized in Table~\ref{tb_stats_band6}.

\textbf{April 16, 2017}\label{04162017}

The Band 6 observations were obtained as part of a five-field mosaic during the time interval 16:00-17:35 UT, with the images synthesized using a restoring beam of approximately 0.6\arcsec\ and mapped onto a grid with a pixel size of 0.1\arcsec. The data were corrected for the primary-beam response of the 12 m antennas out to the half-power point. 

\begin{figure}[!h]
   \centering
    \includegraphics[width=\columnwidth]{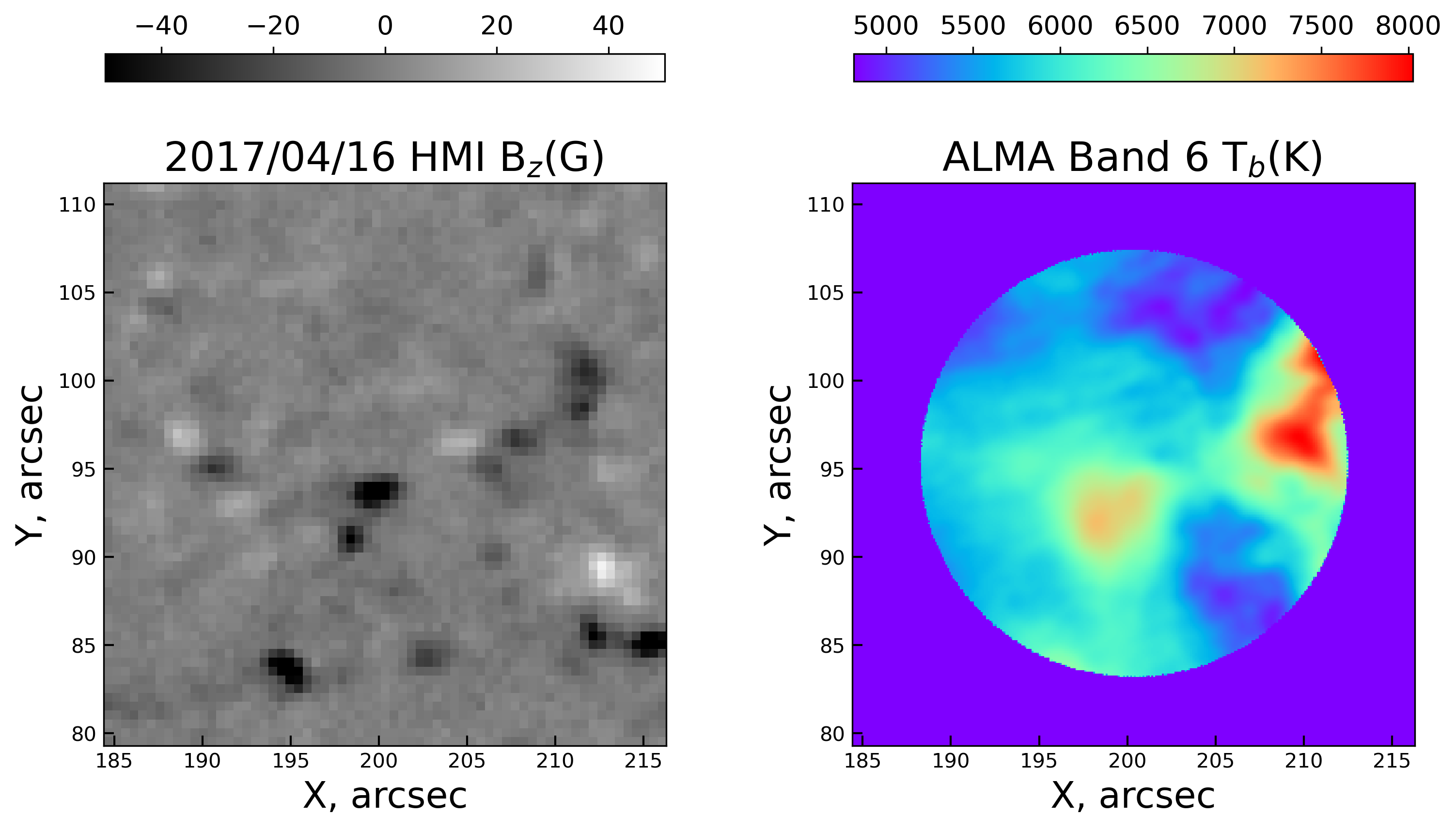}
    \caption{Same as in Fig.~\ref{fig_04272017} for the 2017 April 16 observations in Band 6. }
              \label{fig_04162017}%
    \end{figure}

For the present analysis, we selected one of the mosaic fields containing detectable magnetic structure while still representing very quiet-Sun conditions (Fig.~\ref{fig_04162017}). The HMI magnetogram of this region shows weak and spatially dispersed magnetic elements of both polarities, with vertical magnetic field values ranging from approximately $-94$~G to $+45$~G and a mean unsigned field strength of only about 8.5~G. 

The corresponding ALMA Band 6 brightness temperatures range from approximately 3425 K to 10600 K, with a mean value of about 5975 K and a standard deviation of 895 K (Table~\ref{tb_stats_band6}). The brightest emission is concentrated toward the right side of the field, where the strongest magnetic elements are located. Large areas represent extremely quiet chromospheric conditions with minimal brightness contrast, consistent with the weak underlying magnetic field. These very quiet regions contribute substantially to the low-brightness peak in the temperature distribution.

This data set represents one of the quietest chromospheric environments analyzed in this study. As a result, it is less suitable for direct comparison with the MURaM-ChE snapshot, which represents more strongly magnetized regions. Nevertheless, it provides an important reference for the low-activity regime and illustrates the chromospheric brightness behavior under conditions of minimal magnetic structuring.

\textbf{April 23, 2017}

The Band 6 observations on April 23, 2017 were obtained during the time interval 15:53-16:01 UT and sample the same enhanced-network region as the corresponding Band 3 observations discussed above, although with a smaller field of view and approximately 80 minutes earlier \citep{Molnar2021}. The brightness temperature images were reconstructed from interferometric visibilities using standard ALMA imaging and calibration procedures described in \citet{Molnar2019}, resulting in brightness maps with a pixel size of $0.6\arcsec$.

\begin{figure}[!h]
   \centering
    \includegraphics[width=\columnwidth]{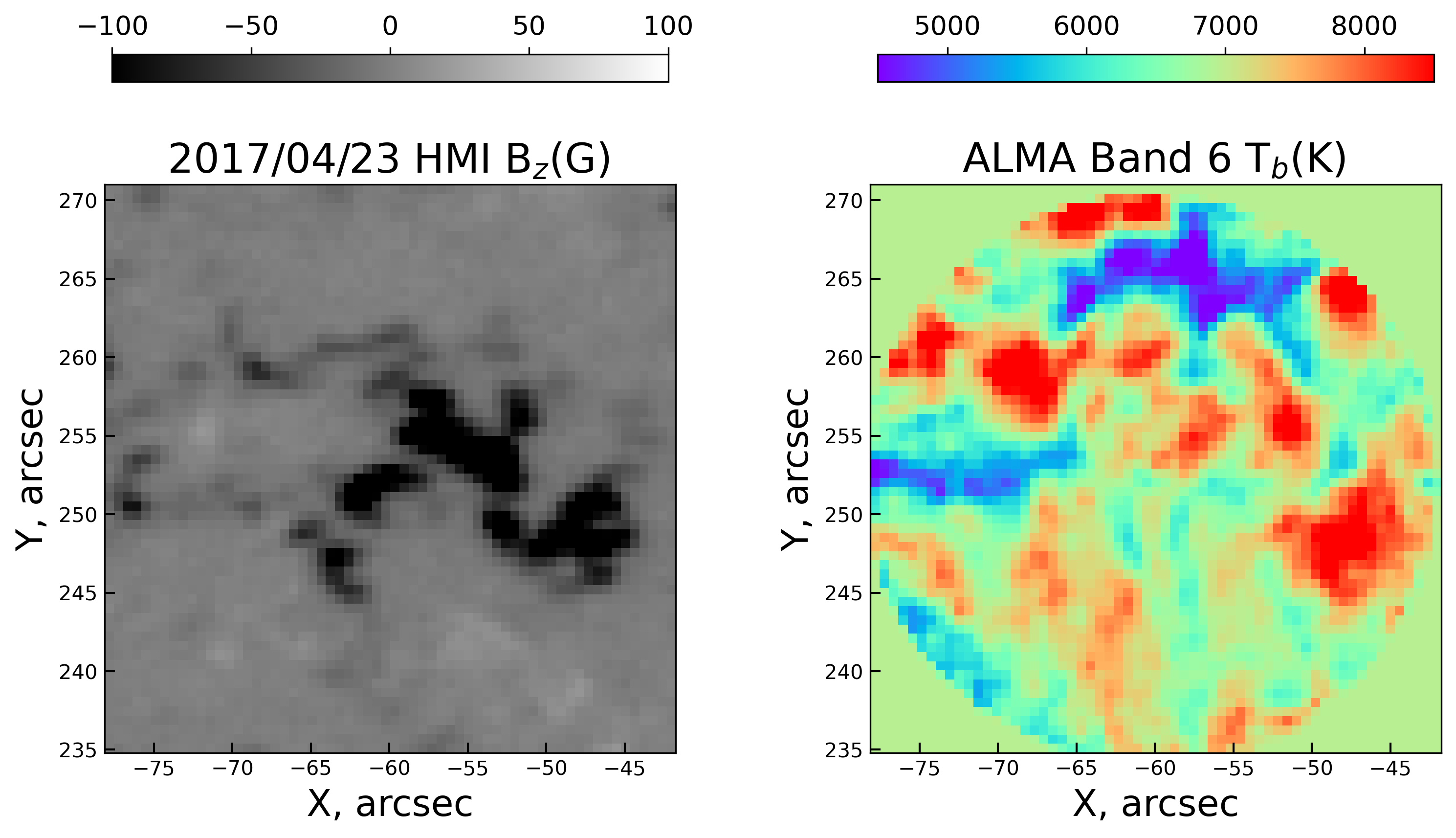}
    \caption{Same as in Fig.~\ref{fig_04272017} for the 2017 April 23 observations in Band 6. }
              \label{fig_04232017_b6}%
    \end{figure}

The HMI magnetogram of the selected 25\arcsec$\times$25\arcsec\ region (Fig.~\ref{fig_04232017_b6}) reveals a compact magnetic configuration dominated by strong negative-polarity concentrations, with smaller positive-polarity elements in close proximity. The mean unsigned photospheric magnetic field strength is approximately 24 G, consistent with ENW conditions.

In contrast to the more extended brightness structures seen in Band 3, the Band 6 emission appears more spatially localized and follows the fine scale magnetic pattern. The brightest regions are concentrated above the strongest magnetic elements, forming relatively compact enhancements. This behavior is consistent with the lower formation height of Band 6 radiation, which samples layers where the magnetic structuring is less spatially expanded (see, e.g. distribution of simulated magnetic field in Fig.~\ref{fig1}b-c).

The brightness temperatures range from approximately 2880 K to 12010 K, with a mean value of about 6955 K and a standard deviation of 965 K (Table~\ref{tb_stats_band6}). This data set exhibits the highest mean brightness temperature and the broadest distribution among the Band 6 observations. 

\begin{table*}
\caption{Statistical properties of the brightness temperature distributions in ALMA Band~6 for the selected observational subregions
and simulations analyzed in this work. The mean, median, standard deviation, minimum, and maximum brightness temperatures are given in K and rounded to the nearest 5~K. }
\label{tb_stats_band6}
\centering
\begin{tabular}{lrrrrrrr}
\hline\hline
Data set & Mean & Median & Std & Min & Max & Skewness & Kurtosis \\
\hline
2017-04-16   & 5975 & 5870 &  895 & 3425 & 10600 &  0.895 &  3.184 \\
2017-04-23   & 6955 & 6995 &  965 & 2880 & 12010 & -0.290 &  1.563 \\
2019-11-09   & 5870 & 5880 &  425 & 4250 &  7280 & -0.076 &  0.591 \\
\hline
MURAM        & 6790 & 6705 & 1200 & 3540 & 10910 &  0.213 & -0.389 \\
Bifrost 385  & 5000 & 4880 &  800 & 3255 &  9565 &  0.964 &  1.458 \\
Bifrost 463  & 4595 & 4465 &  775 & 3050 &  9310 &  1.399 &  3.196 \\
\hline

\end{tabular}
\end{table*}

\textbf{November 09, 2019}

The Band 6 observations on November 9, 2019 were obtained as a single-field pointing during the time interval 13:51-14:35 UT, earlier in the day than the corresponding Band 3 observations discussed above. The images were mapped onto a grid with a pixel size of 0.1\arcsec\ and restored with a synthesized beam of approximately 0.6\arcsec. The data were corrected using the 12 m antenna primary beam out to the half-power point. As in the Band 3 observations of this date, the exact pointing location is not precisely known, which limits the accuracy of coalignment with other instruments and complicates detailed comparisons with magnetic field measurements.

The selected field represents predominantly quiet-Sun conditions, with only weak and sparsely distributed magnetic elements present within the field of view (unsigned photospheric magnetic field does not exceed 10 G). The absence of strong magnetic concentrations results in relatively homogeneous chromospheric brightness, with only minor localized enhancements associated with small-scale magnetic features. The brightness temperature ranges from approximately 4250 K to 7280 K, with a mean value of about 5870 K and a standard deviation of 425 K. Compared to the 23 April 2017 Band 6 observations, this data set shows both lower average brightness and reduced variability, consistent with very quiet chromospheric conditions.

\subsection{Comparison with ALMA observations}\label{comparison}
\subsubsection{Relationship between brightness temperature and magnetic field strength}
\begin{figure}
   \centering

   \includegraphics[width=\columnwidth]{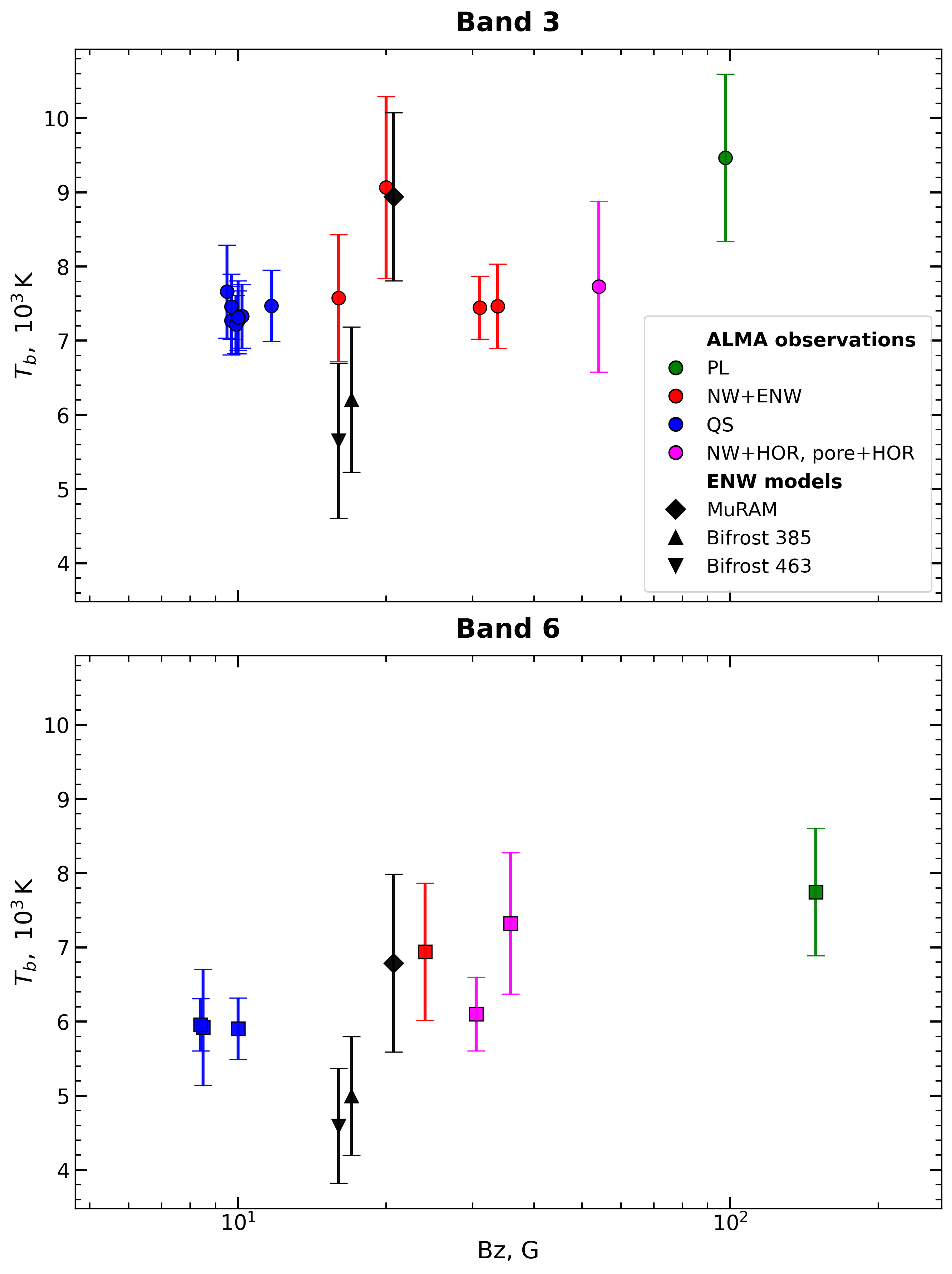}
   \caption{Brightness temperature as a function of vertical magnetic field strength $|B_z|$ for ALMA Bands 3 (top) and 6 (bottom). Colored symbols show ALMA observations for selected solar features: plage (PL, green), network and enhanced network (NW+ENW, red), quiet Sun (QS, blue), and network or pore with pronounced horizontal fields (NW+HOR, pore+HOR,  magenta). Black symbols represent ENW model predictions from numerical simulations: MURaM-ChE (diamond), Bifrost 385 (upward triangle), and Bifrost 463 (downward triangle). 
   }
              \label{fig_tb_mf}%
\end{figure}

Figure~\ref{fig_tb_mf} shows the relationship between brightness temperature and the unsigned vertical magnetic field strength $|B_z|$ for ALMA observations in Bands~3 and 6, taken from Table~\ref{table_observations}, together with predictions from the ENW numerical models, including the MURaM-ChE, Bifrost 385, and Bifrost 463 snapshots. In both bands, the observations demonstrate a clear overall trend of increasing brightness temperature with increasing magnetic field strength. Quiet-Sun regions (QS) with weak magnetic fields ($|B_z| \lesssim 10$~G) exhibit the lowest brightness temperatures, while network and enhanced-network regions (NW-ENW) with intermediate magnetic field strengths show systematically higher values. The highest brightness temperatures are observed in plage regions, where strong magnetic concentrations produce substantial chromospheric heating.

In Band~3, QS observations cluster around brightness temperatures of approximately 7200–7600~K, while NW–ENW regions reach values of about 7400-9100~K. The plage observation stands out with significantly higher brightness temperatures near 9500~K. In Band~6,  QS regions exhibit brightness temperatures near 5800-6000~K, NW-ENW regions reach approximately 6100-7000~K, and plage regions attain the highest values of about 7700~K.

The ENW model predictions follow the same general trend. The MURaM-ChE snapshot produces brightness temperatures that lie within the range of NW–ENW observations in both bands, indicating good agreement with the observed brightness in moderately magnetized chromospheric regions. The Bifrost snapshots produce systematically lower brightness temperatures for both NW-ENW and QS magnetic field strengths. 
This difference is consistent across both bands and likely reflects differences in atmospheric density stratification and the corresponding formation heights of millimeter emission. Since the brightness temperature reflects the local temperature at the height where the emission becomes optically thick, lower atmospheric densities in the Bifrost simulations would shift the effective formation height downward to cooler layers, resulting in lower brightness temperatures even in the presence of comparable magnetic field strengths. This illustrates the value of comparing chromospheric models with ALMA observations. According to \citet{Loukitcheva2015}, the effective formation heights for Bands~6 and 3 in the Bifrost snapshot 385 are about 900~km and 1500~km, respectively, which is considerably lower than the heights of $1900\pm550$~km and $2700\pm850$~km estimated from the MURaM-ChE snapshot (Table~\ref{table1}).

The observations corresponding to regions with NW or pore morphology and additionally exhibiting strong horizontal magnetic fields (+HOR), identified in the 3D extrapolations of the magnetic field by \citet{Jafarzadeh2021}, demonstrate brightness temperatures generally comparable to typical NW-ENW regions, but with a tendency toward lower brightness than expected from the magnetic field strength. This behavior highlights the importance of the full magnetic field topology in determining chromospheric thermal structure. Regions with strong horizontal fields introduce additional scatter into the brightness–magnetic field relationship and weaken the direct correspondence between brightness temperature and $|B_z|$ alone.

Despite this scatter, the overall agreement between the MURaM-ChE model and the NW-ENW observations in both bands indicates that the model reproduces the thermal and magnetic conditions of these regions reasonably well. The model brightness temperatures and magnetic field strengths occupy the same parameter space as the ENW observations, supporting the interpretation that the MURaM-ChE snapshot provides a realistic representation of chromospheric conditions in moderately magnetized regions, although its brightness temperatures lie toward the upper end of the observed range.

\subsubsection{Brightness temperature distributions}

\begin{figure}
   \centering
    \includegraphics[width=\columnwidth]{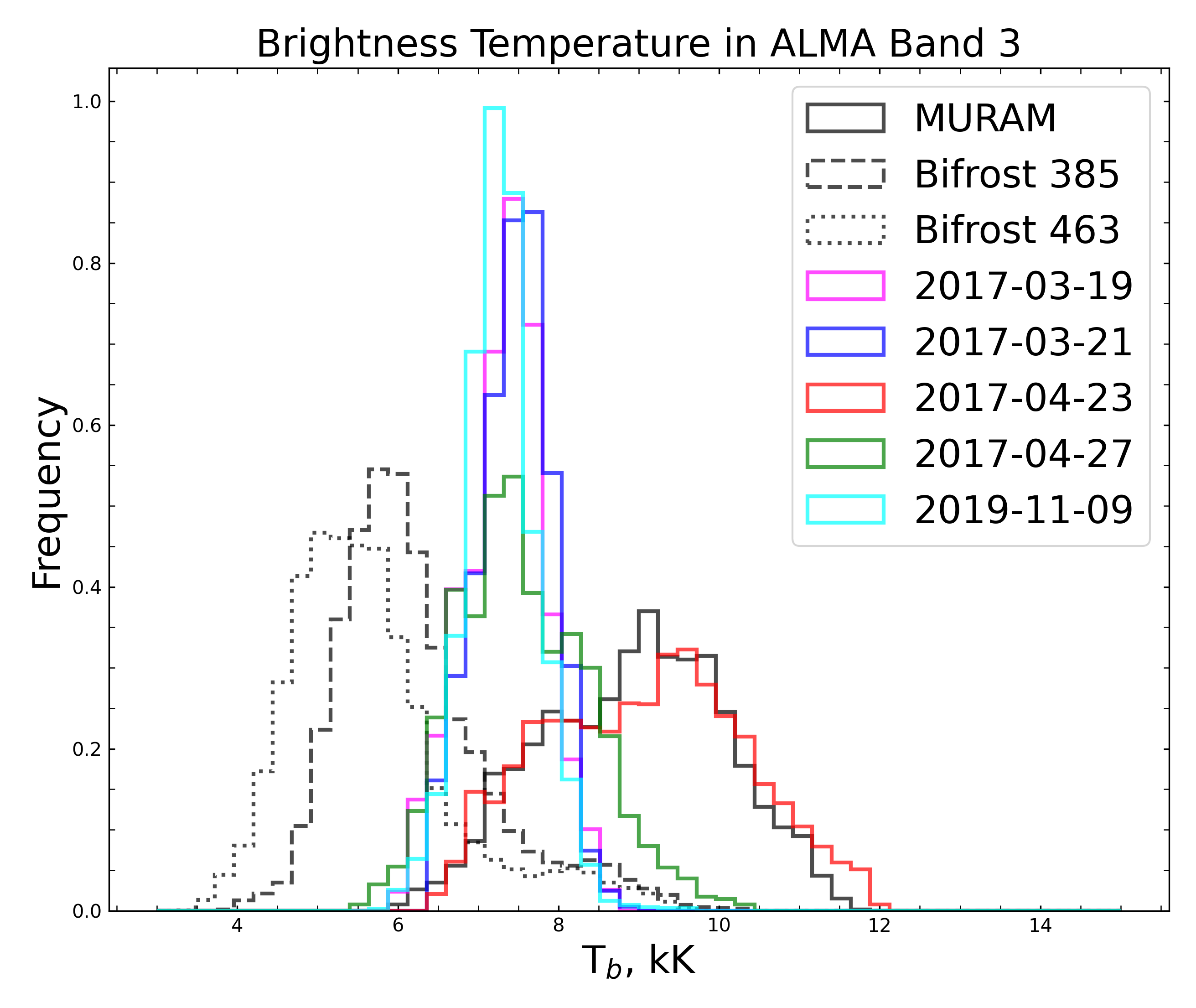}
    \caption{Brightness temperature distributions in ALMA Band 3, each normalized independently to unit area. Black curves show results from MURaM-ChE (solid),  Bifrost 385 (dashed), and Bifrost 463 (dotted) simulations, while colored curves represent ALMA observations from different observing dates. }
              \label{fig_hist3}%
    \end{figure}

\begin{figure}
   \centering
    \includegraphics[width=\columnwidth]{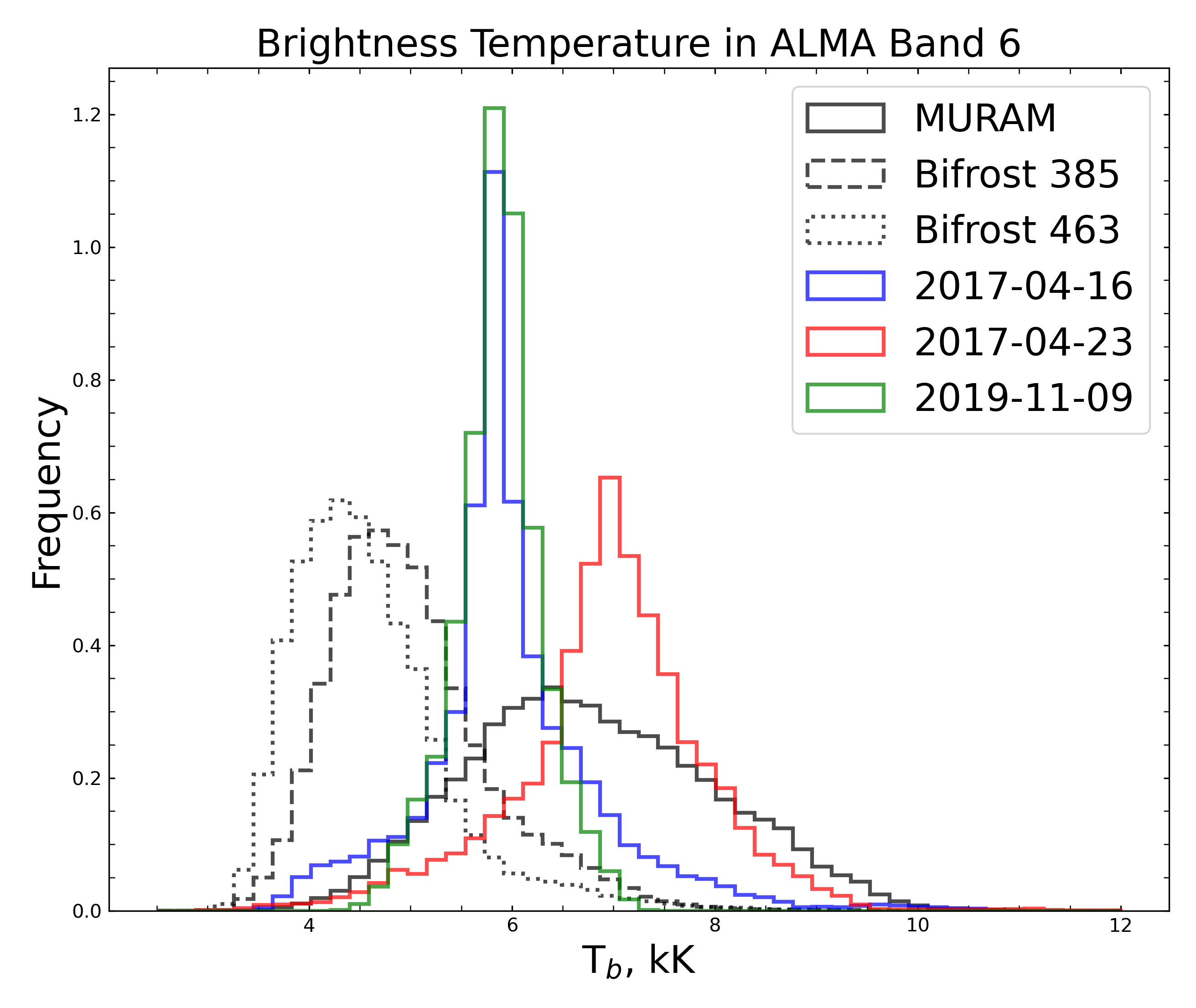}
    \caption{Same as in Fig.~\ref{fig_hist3} for ALMA Band 6. }
              \label{fig_hist6}%
    \end{figure}

    Figures~\ref{fig_hist3} and ~\ref{fig_hist6} compare the one-dimensional brightness temperature distributions derived from the MURaM-ChE and Bifrost simulations with  the ALMA observations in Bands 3 and 6, described in Sects.~\ref{band3} and ~\ref{band6}, respectively.  In Band 3, the degraded MURaM-ChE snapshot produces a brightness distribution that is clearly shifted toward higher temperatures, with a mean value of about 8900~K and a peak near $\sim$9000~K. In contrast, both Bifrost snapshots are systematically shifted to significantly lower temperatures, with maxima near 6000~K and 5600~K, respectively. The separation between the MURaM and Bifrost distributions is substantial, with only limited overlap, indicating markedly different thermal regimes in the two models. A similar contrast is observed in Band 6. The MURaM-ChE distribution peaks near 6700~K, whereas the Bifrost distributions remain centered at much lower temperatures, around 5000~K and 4600~K. While the Bifrost distributions exhibit broader and more asymmetric shapes, their extended tails remain confined to lower temperature ranges. 
    
Overall, the comparison shows that the dominant difference between the models is a systematic temperature offset, with the MURaM-ChE snapshot consistently producing higher brightness temperatures across both ALMA bands, while differences in the shape of the distributions are also present, particularly in the degree of asymmetry and width. As discussed above, these differences likely reflect variations in the thermal and density structure of the simulated atmospheres. The lower brightness temperatures in the Bifrost simulations are consistent with emission forming, on average, in cooler atmospheric layers. This behavior is observed in both Bands 3 and 6, suggesting that the underlying differences between the models affect the formation of millimeter emission over a range of chromospheric heights, rather than being confined to a specific layer.

The observed brightness distributions exhibit a range of shapes and peak positions reflecting the diversity of magnetic environments sampled. In Band 3, the 23 April 2017 data set shows the closest agreement with the MURaM-ChE model. Its histogram peaks near 9000~K and displays a broad distribution with an extended high-temperature tail, closely resembling the shape and width of the MURaM-ChE distribution. In contrast, the observations from 19 March, 21 March, 27 April 2017, and 9 November 2019, exhibit narrower distributions with peaks near 7000- 8000 K and reduced high-temperature tails. These distributions remain significantly warmer than the Bifrost predictions but do not reach the highest temperatures present in the MURaM-ChE model, indicating intermediate chromospheric conditions.

A similar relationship between the models and observations is found in Band 6. The 23 April 2017 data set again provides the best agreement with the MURaM-ChE model, with a histogram peak near 7000~K and a broad temperature range comparable to the simulated distribution. This observation samples the same magnetically structured region as the Band 3 data and exhibits enhanced brightness associated with concentrated magnetic flux. The 16 April 2017 and 9 November 2019 observations, which correspond to very quiet-Sun conditions, display significantly narrower distributions centered near 6000~K with limited high-temperature tails. These distributions are closer to the lower-temperature regime predicted by the Bifrost models but remain slightly warmer and broader.

Taken together, the Band 3 and Band 6 comparisons show that the observed brightness distributions for the main part lie between the predictions of the MURaM-ChE and Bifrost models. The MURaM-ChE model successfully reproduces both the location of the histogram maximum and the extent of the high-temperature tail observed in magnetically active quiet-Sun regions, particularly in the 23 April 2017 data set. In contrast, the systematically lower and narrower Bifrost distributions indicate formation of simulated mm emission in cooler atmospheric layers. 

\subsubsection{Spatially-averaged millimeter brightness spectrum}

\begin{figure}
   \centering
   \includegraphics[width=\columnwidth]{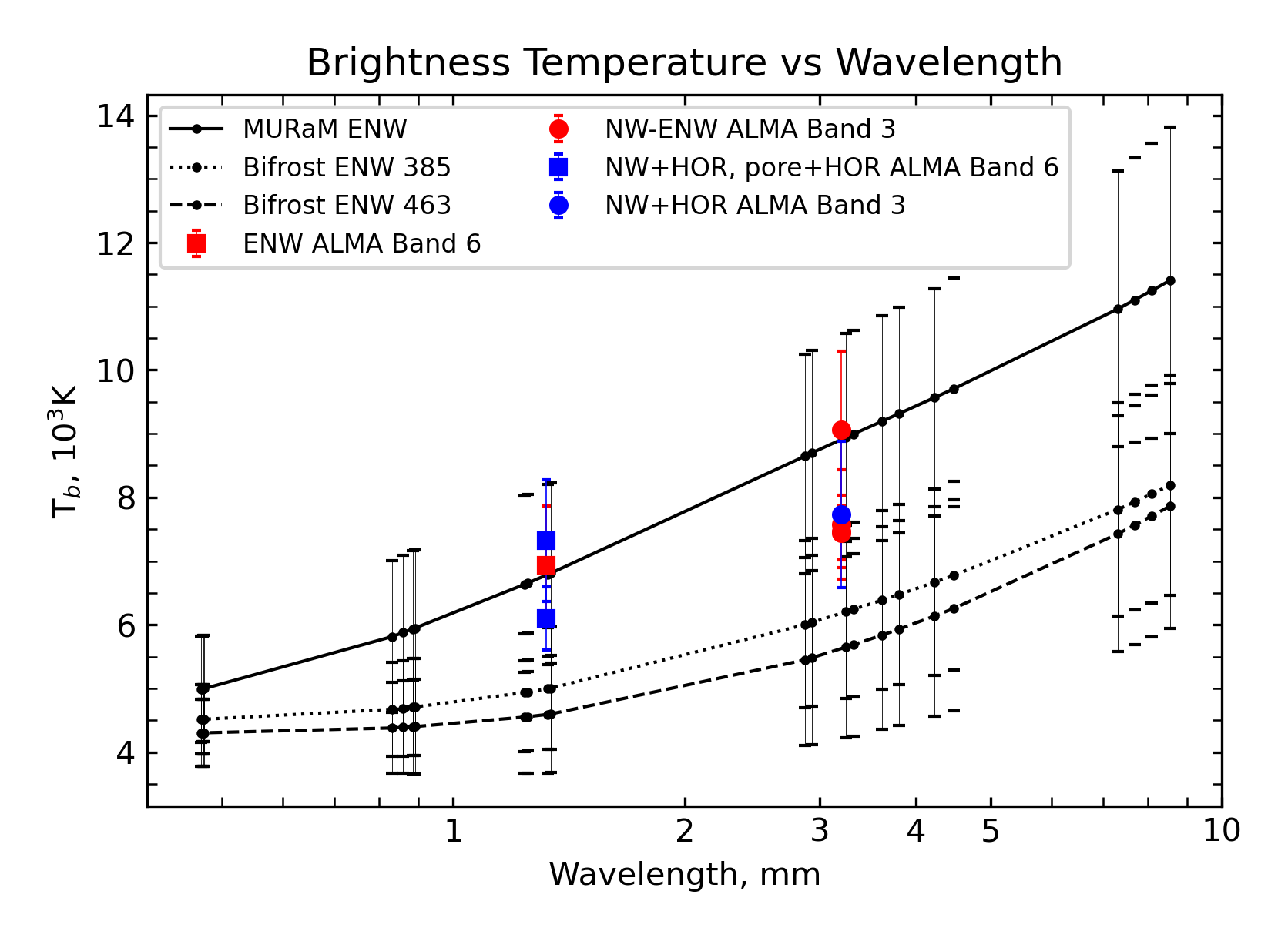}
   \caption{Spatially-averaged brightness temperature as a function of mm wavelength. Black curves show the distributions predicted by ENW numerical models as a function of wavelength: MURaM-ChE (solid), Bifrost 385 (dotted), and Bifrost 463 (dashed), with vertical bars indicating the model spread. Colored symbols denote ALMA observational values at two wavelengths corresponding to Bands 3 and 6: NW-ENW measurements (red), and NW+HOR and pore+HOR measurements (blue), shown as circles for Band 3 and squares for Band 6.}
              \label{fig_tb_lambda}%
\end{figure}

For a broader perspective on the behavior of the mm brightness, Fig.~\ref{fig_tb_lambda} shows the spatially averaged brightness temperature as a function of wavelength for the ENW numerical models, together with ALMA measurements of network and enhanced-network regions (NW-ENW) and network and pore regions with strong horizontal fields (HOR) in Bands~6 and~3.

All three models predict a monotonic increase of brightness temperature with wavelength, reflecting the shift of the effective formation height toward progressively higher chromospheric layers where the temperature is, on average, higher. However, the absolute brightness levels and the rate of increase differ significantly between the models.  The MURaM-ChE model predicts systematically higher brightness temperatures and a steeper increase with wavelength than either Bifrost simulation. This behavior indicates differences in the thermal
and density stratification of the model atmospheres that become more
pronounced with increasing formation height, mainly in the lower
and middle chromosphere.

In the MURaM model, the average density at typical chromospheric temperatures is approximately an order of magnitude higher than in the Bifrost models \citep{Ondratschek2024}. Identifying the cause of this difference is challenging. The chromospheric density scale height depends on the magnetic field, turbulent motions, and local heating and cooling. Two differences between the models stand out: MURaM has approximately twice the photospheric RMS magnetic field strength and significantly higher chromospheric kinetic energy. A broader set of simulations would be needed to determine whether either difference is the dominant cause of the density enhancement.

The ALMA measurements follow the same overall trend. The NW-ENW observations lie between the Bifrost and MURaM-ChE predictions at both wavelengths. In Band~6, the only one available ENW observational point lies above the Bifrost values and very close to the MURaM-ChE prediction. In Band~3, three NW-ENW measurements are available, with the highest Band~3 value showing a particularly good agreement with MURaM-ChE and two others lying in between the two models within the distribution spread.

The NW+HOR and pore+HOR measurements also follow the general increase of brightness with wavelength. The apparent spread again mainly reflects the limited number of available observations. In Band~6, the values span the range around the simulated MURaM-ChE brightness, whereas in Band~3 the NW+HOR measurement occupies intermediate levels between the two models. These results are consistent with the more complex magnetic topology of NW+HOR and pore+HOR regions, but the small number of available data sets does not allow stronger statistical conclusions.

Overall, the observational measurements support the general wavelength dependence predicted by the numerical simulations while also showing that the observed chromospheric conditions span a range of thermal states. The NW-ENW observations are broadly consistent with values between the Bifrost and MURaM-ChE predictions, with the closest agreement with MURaM-ChE found for the brightest ENW case of 23 April 2017 in both bands.

\section{Conclusions and Outlook}

We tested the MURaM-ChE rMHD model against ALMA
millimeter-continuum observations of the solar chromosphere in
Bands~3 and 6 and compared its predictions with those of Bifrost
ENW simulations. Because millimeter brightness temperature
closely traces the local gas temperature near optical depth unity
\citep[e.g.,][]{Loukitcheva2004,Loukitcheva2009,Wedemeyer2016},
these observations provide a direct diagnostic of chromospheric
thermal structure.

The MURaM-ChE model reproduces the main properties of the
observed ENW brightness distributions, including their mean
values, widths, and overall shapes. The closest agreement is found
for the 23 April 2017 data set in both bands, including the
extended high-temperature tail. Although the simulation has a
bipolar magnetic configuration and the observed region is
dominated by a single polarity, their agreement indicates that
the millimeter brightness is determined primarily by local
magnetic and thermodynamic conditions rather than by the
large-scale polarity structure.

The level of agreement varies with the magnetic environment.
QS regions generally show narrower brightness distributions and
lower contrast, whereas more strongly or complexly magnetized
regions exhibit broader distributions and enhanced
high-temperature tails. Some regions classified as ENW
nevertheless have QS-like distributions, possibly because of
limited spatial sampling, instrumental resolution, or differences
in atmospheric conditions.

The Bifrost simulations produce systematically lower brightness
temperatures than MURaM-ChE in both ALMA bands, with distributions
shifted below most ENW and even some QS observations. This difference
can be attributed, at least partly, to the lower atmospheric densities
in the Bifrost models, which shift the effective formation heights of
the optically thick millimeter emission toward deeper and cooler layers.
The average unsigned magnetic field strengths of the Bifrost snapshots,
approximately 16--17~G after degradation to HMI resolutions, are also somewhat lower than in the MURaM-ChE
snapshot and may further contribute to the difference in brightness.

Both observations and simulations show an overall increase in
brightness temperature with vertical magnetic field strength,
consistent with earlier results \citep[e.g.,][]{Loukitcheva2009}.
The substantial scatter, particularly in regions with strong
horizontal fields, indicates that the emission depends on the
full magnetic structure rather than on $|B_z|$ alone.

Overall, the MURaM-ChE simulation provides a physically
consistent representation of ENW chromospheric conditions and
captures the principal properties of the observed millimeter
emission. However, the variability among observed regions,
effects of spatial resolution and sampling, and limited number
of Band~6 data sets restrict the extent to which these conclusions
can be generalized.

Future work will include a spatially resolved comparison with
the 23 April 2017 observations, analysis of the multi-peaked
contribution functions and effective formation heights, and
time-dependent studies of shocks and evolving magnetic
structures. These investigations will clarify how local
thermodynamic and magnetic conditions shape the millimeter
brightness and will provide stronger constraints on
chromospheric energy balance and structure.

\begin{acknowledgements}
      We thank Momchil E. Molnar and collaborators for providing the ALMA Band~3 and Band~6 brightness-temperature maps originally presented and analyzed by \citet{Molnar2019,Molnar2021}, and for permitting their use in this study.  
     This paper makes use of the following ALMA data: ADS/JAO.ALMA\#2016.1.00423.S, ADS/JAO.ALMA\#2016.1.00572.S, ADS/JAO.ALMA\#2016.1.00030.S,  ADS/JAO.ALMA\#2016.1.00788.S, ADS/JAO.ALMA\#2016.1.00050.S, ADS/JAO.ALMA\#2016.1.01129.S, ADS/JAO.ALMA\#2016.1.01532.S, ADS/JAO.ALMA\#2016.1.00202.S, ADS/JAO.ALMA\#2017.1.00653.S, ADS/JAO.ALMA\#2019.1.00516.S, ADS/JAO.ALMA\#2017.1.01672.S. ALMA is a partnership of ESO (representing its member states), NSF (USA) and NINS (Japan), together with NRC (Canada), NSTC and ASIAA (Taiwan), and KASI (Republic of Korea), in cooperation with the Republic of Chile. The Joint ALMA Observatory is operated by ESO, AUI/NRAO and NAOJ.
      This project has received funding from the European Research Council (ERC) under the European Union's Horizon 2020 research and innovation programme (grant agreement No. 101097844 — project WINSUN).  We are grateful for the computational resources provided by the Raven and Viper supercomputer systems of the Max Planck Computing and Data Facility (MPCDF) in Garching, Germany. The work of D.P. was funded by the Federal Ministry for Economic Affairs and Climate Action (BMWK) through the German Space Agency at DLR based on a decision of the German Bundestag (Funding code: 50OU2201). 
\end{acknowledgements}

\bibliographystyle{aa}
\bibliography{Loukitcheva2026}

\end{document}